\documentclass[12pt]{iopart}
\usepackage{graphicx}

\usepackage{hyperref}
\usepackage{amstext}  
\usepackage{amssymb}

\newcommand{\FSR}{\textit{FSR}}
\newcommand{\Fi}{\mathcal{F}}
\newcommand{\clipfac}{\alpha_{\textrm{cl}}}
\newcommand{\mmeff}{\epsilon} 
\newcommand{\mmeffmin}{\epsilon_{\min}}
\newcommand{\fceff}{{\eta_{f}}} 
\newcommand{\reff}{\eta_{r}} 
\renewcommand{\L}{\mathcal{L}}
\renewcommand{\S}{\mathcal{S}}
\newcommand{\A}{\mathcal{A}}
\newcommand{\R}{\mathcal{R}}
\newcommand{\T}{\mathcal{T}}
\newcommand{\LTot}{\L_{\textrm{tot}}}
\newcommand{\LCl}{\L_{\textrm{cl}}}

\begin{document}

\title{Fiber Fabry-Perot cavity with high finesse}

\author{D.~Hunger$^2$, T.~Steinmetz$^{1,2,\dagger}$, Y.~Colombe$^{1,\ddagger}$,
  C.~Deutsch$^1$, T.~W.~H{\"a}nsch$^2$ and J.~Reichel$^1$}
\address{$^1$Laboratoire Kastler Brossel, ENS/UPMC-Paris 6/CNRS, 24
  rue Lhomond, F-75005 Paris, France}
\address{$^2$Max-Planck-Institut f{\"u}r
  Quantenoptik/Ludwig-Maximilians-Universit\"at,
  Schellingstra{\ss}e~4, D~80799~M{\"u}nchen, Germany}
\address{$^\dagger$Present address:
    Menlo Systems GmbH, D-82152 Martinsried (Munich), Germany}
\address{$^\ddagger$Present address:NIST, Boulder, Colorado 80305, USA}
\ead{jakob.reichel@ens.fr}



\begin{abstract}
  We have realized a fiber-based Fabry-Perot cavity with CO$_2$
  laser-machined mirrors. It combines very small size, high finesse
  $\Fi\ge130000$, small waist and mode volume, and good mode matching
  between the fiber and cavity modes. This combination of features is
  a major advance for cavity quantum electrodynamics (CQED), as shown
  in recent CQED experiments with Bose-Einstein condensates enabled by
  this cavity [Y. Colombe et al., Nature \textbf{450}, 272 (2007)].
  It should also be suitable for a wide range of other applications,
  including coupling to solid-state emitters, gas detection at the
  single-particle level, fiber-coupled single-photon sources and
  high-resolution optical filters with large stopband.
\end{abstract}

\pacs{07.60.Vg, 42.50.Ex, 42.50.Pq, 42.81.Wg}

\section{Introduction}

Cavity quantum electrodynamics (CQED) experiments provide insight into
the fundamental concepts of quantum mechanics such as entanglement,
decoherence and measurement \cite{Haroche06,Mabuchi02}.  Additionally,
optical CQED is currently growing into a key role in quantum
information processing \cite{Zoller05}, where the transmission of
quantum states between distant nodes is a central problem
\cite{Kimble08}, and in entanglement-enhanced metrology
\cite{Banaszek09}, where the use of optical cavities has very recently
lead to convincing demonstrations of metrologically useful spin
squeezing capable of improving the stability of atomic clocks
\cite{Leroux10}.  A fundamental figure of merit of the
atom-cavity system is its cooperativity $C$, which up to factors of
order 1 is given by
\begin{equation}
  C\sim \frac{\sigma_0}{\pi w_0^2}\Fi\,,
\end{equation}
where $\pi w_0^2$ it the cross-section and $\Fi$ the finesse of the
cavity mode, and $\sigma_0$ is the (effective) scattering
cross-section of the emitter(s) placed in this mode. For example, the
Purcell effect arises when $C>1$ and the efficiency of an important
class of single-photon sources scales as $C/C+1$ \cite{Law97}.  In
addition to having a high ccoperativity, cavities used in quantum
information applications should be miniaturized and fiber-coupled so
that they can be used in scalable setups.  This motivates considerable
efforts to develop miniature high-finesse cavities \cite{Vahala03}.
However, no cavity type so far unites all desired properties in a
single device.  Therefore, progress in CQED and its applications
hinges on the development of new, miniature cavities with high
cooperativity.

Here we describe a fiber-based Fabry-Perot cavity (FFPC) that combines
tunability and high finesse with excellent and stable coupling to
single-mode optical fibers, which is achieved without mode-matching
optics. The waist $w_0$ is about an order of magnitude smaller than in
the macroscopic high-finesse cavities typically used in optical CQED
experiments \cite{Haroche06,Mabuchi02}. The cavity mode is located in
free space, making it easy to couple to atomic emitters. The cavity
design is based on a new laser machining process where a single,
focused CO$_2$ laser pulse creates a concave depression in the cleaved
fiber surface.  The first use of this FFPC was in an experiment that
demonstrated strong-coupling CQED with Bose-Einstein
condensates on atom chips \cite{Colombe07}, Prompted by these results,
the use of such cavities is now being explored with trapped neutral
atoms and ions \cite{Luo09,Kimble08}, but also with color centers in
diamond \cite{Jelezko06}, semiconductors \cite{Deveaud06,Shields07}
and vibrating sub-wavelength objects \cite{Jayich08,Favero09}. In all
of these applications, the small size, ruggedness and built-in fiber
coupling of the FFPC are advantageous, favouring its use in hostile
environments such as cryostats.
Moreover, we have already obtained single-atom cooperativities $C>100$
\cite{Colombe07}, and still higher values can be expected when using
state-of-the art dielectric coatings. This combines with the good
fiber coupling, such that decisive progress can be expected from this
FFPC for several applications: It should lead to single-photon sources
\cite{Lounis05} with exceptional overall performance, where not only
an atomic excitation is converted into a cavity photon with a
probability close to one, but also this photon is efficiently
extracted into a single-mode optical fiber. For similar reasons, the
FFPC may improve the performance of quantum memories
\cite{Luo09,Kimble08}, where currently the overall conversion
efficiency between the memory qubit and the desired optical mode is
still low. It will also enable single-atom state detection (``qubit
readout'') with high quantum efficiency, and may enable optical
detection with less than a single photon scattered by the emitter
\cite{Poldy08}.  In this article, we present a detailed theoretical
and experimental investigation of this new type of fiber cavity.

\section{Principle of the fiber Fabry-Perot cavity}

\begin{figure}[htb]
  \centering
  \includegraphics[width=0.6 \textwidth]{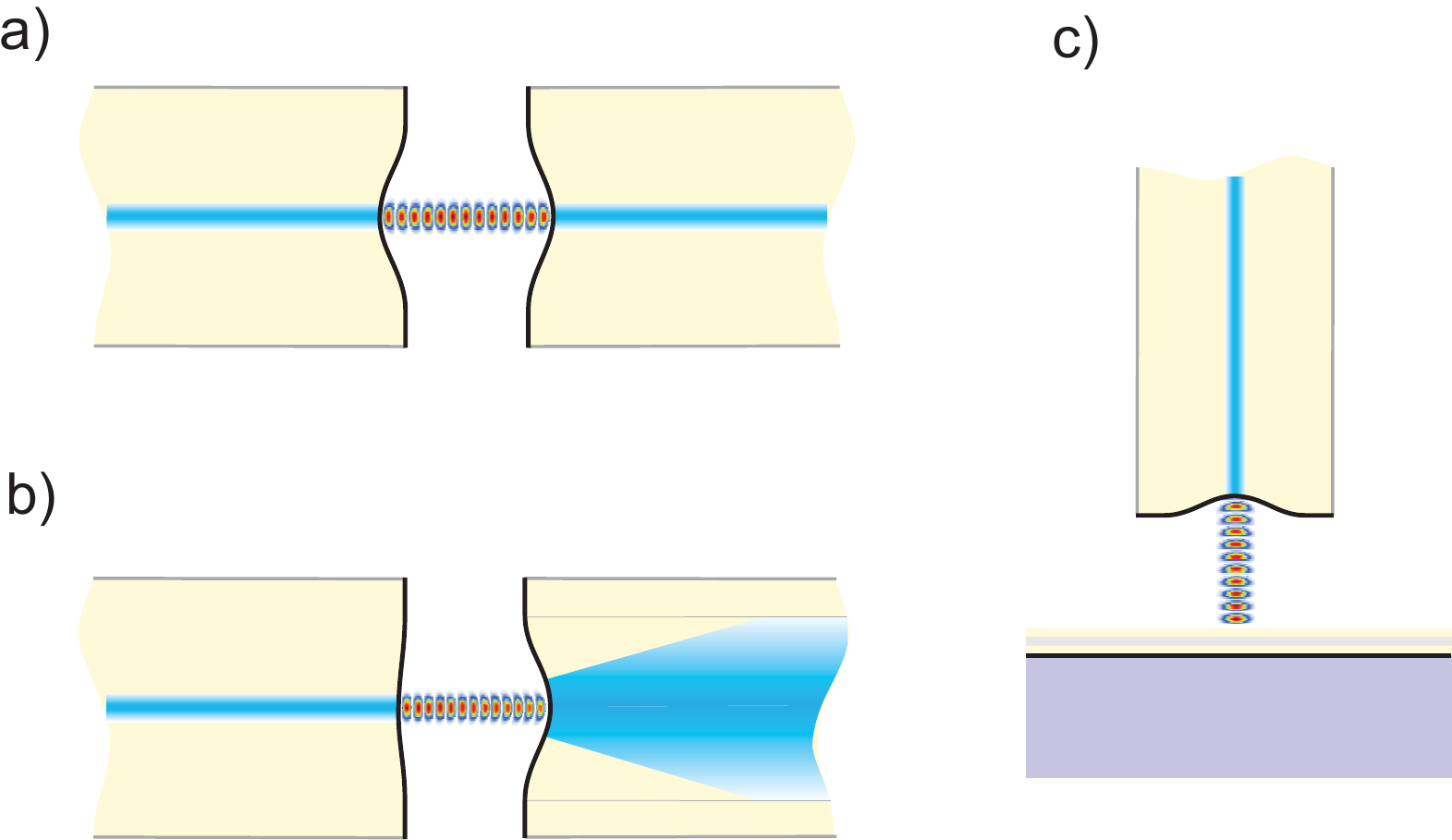}
  \caption{Possible cavity geometries. (a) Cavity made from two
    single-mode fibers; (b) one single-mode and one multi-mode fiber;
    (c) single fiber perpendicular to a reflective planar
    substrate. \label{fig:scheme} }
\end{figure}

The core of our cavity design is a concave, ultralow-roughness mirror
surface fabricated directly on the end face of an optical fiber.
FFPCs with mirrors on fiber end faces have been built before
\cite{Salik02, Steinmetz06,Trupke05,Muller09}, but until now had only
moderate finesse (up to a few 1000), limited by the methods used to
fabricate the concave surface on the fiber. Here we use a single
CO$_2$ laser pulse to shape the concave mirror surface, which is then
coated with a high-performance dielectric coating. As we show below,
this improves the finesse by more than an order of magnitude over the
older methods and gives access to an interesting range of small radius
of curvature (ROC).  As in our earlier design using glued mirrors
\cite{Steinmetz06}, a stable cavity is then constructed either from
one mirror fiber tip facing a macroscopic, planar or concave mirror,
or from two closely spaced fiber tips placed face to face
(fig.~\ref{fig:scheme}). In this article, we will be mainly interested
in the second variant.

Radii of curvature can be fabricated down to $50\,\mu$m and probably
below.  Because of the small mirror diameter $D$ (smaller than the
fiber diameter, which is typically $125\,\mu$m), the mirrors can
approach each other very closely (down to a few $\lambda/2$) without
touching. The result is a very small mode waist $w_0$ between 1 and
$2\,\mu$m, and a small mode volume down to a few $\lambda^3$.  The
small $w_0$ combined with a length much shorter than the Rayleigh
range is the reason for the excellent SM fiber coupling efficiency
with no need for mode-matching optics.  as we discuss theoretically in
sec.~\ref{sec:coupling} and demonstrate experimentally in
sec.~\ref{sec:couplingExp}.

\section{Fabrication of concave mirrors on optical fibers}

\subsection{Laser machining}
\label{sec:Lasermachining}

\begin{figure}[htb]
  \centering
  \includegraphics[width=0.75 \textwidth]{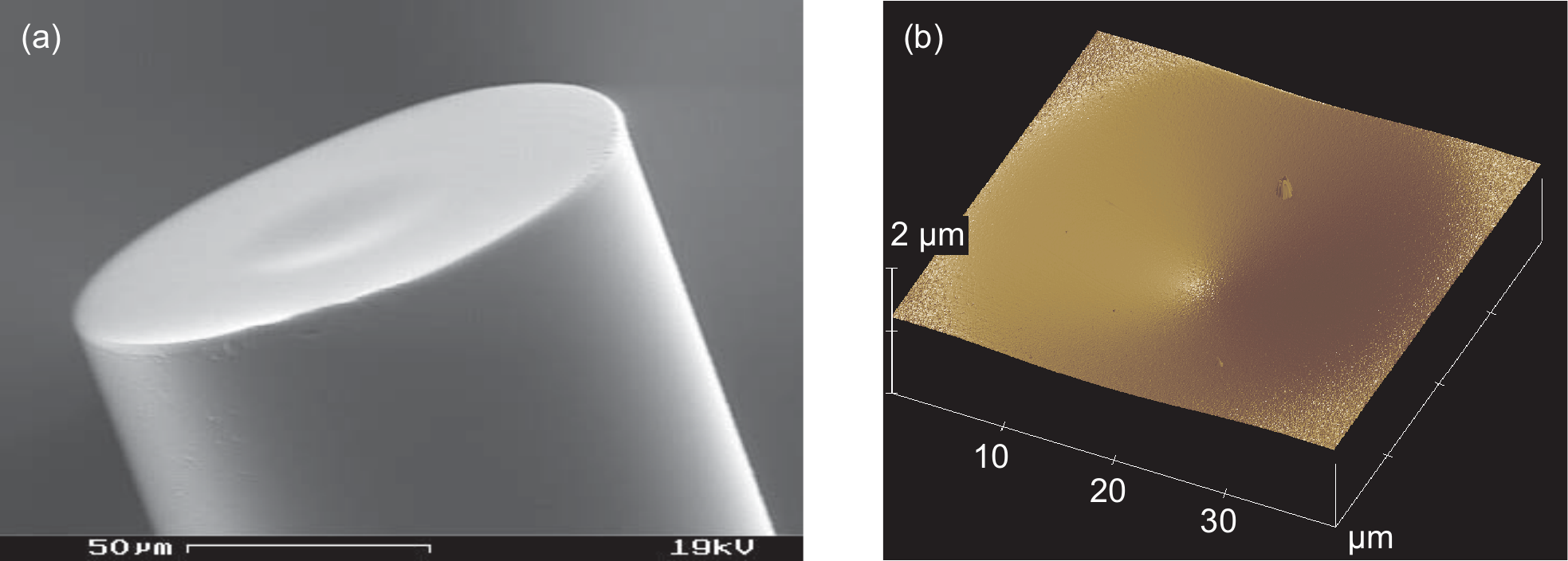}
  \caption{(a) SEM image of a laser-machined fiber end face. 
    (b) AFM measurement of a processed fiber end face with a
    single pulse of $1.03\,$W and pulse length $40\,$ms, focused to
    $w_0=24\,\mu$m.
    \label{fig:surface}}
\end{figure}

As will be described in more detail in \cite{Hunger08b}, we have found
a parameter regime where a single pulse of CO$_2$ laser radiation
focussed onto the cleaved end face of an optical fiber produces a
concave surface with extremely low roughness
(figure~\ref{fig:surface}). In this regime, thermal evaporation
occurs, while melting is restricted to a thin surface region, avoiding
global contraction into a convex shape. This sets it apart from the
regimes used in CO$_2$ laser-based fabrication of microspheres
\cite{Vernooy98} and transformation of microdiscs into high-Q
microtoroid resonators \cite{Armani03}.

We can currently laser machine structures with ROC
between $40\,\mu$m and 2\,mm, diameters between 10 and $45\,\mu$m, and
a surface roughness of about 0.2\,nm rms in the optical range. These
structures are obtained with CO$_2$ waist sizes between 18 and
$80\,\mu$m, powers in the range $P=0.3\ldots 1.1\,$W and pulse lengths
between 5 and 400\,ms.  A dichroic beam splitter and optical
microscope are used to align the fiber with respect to the CO$_2$
laser focus. In our current setup, we estimate the lateral alignment
precision to be better than 2$\,\mu$m.

\subsection{Surface analysis}
\label{sec:surface}

We have laser machined a large number ($>500$) of fiber mirror
structures. Initially, we used an AFM\footnote{Digital Intrsuments
  Dimension 3100} to get both profile and roughness data. Scanning the
full surface to determine curvatures is a slow process however, so
that we could only measure a small number of fibers with this
method. To get approximate values for other fibers, we compared them
to the well-characterized ones under an optical microscope. More
recently, we have used an interferometric microscope\footnote{Fogale
  Micromap3D} for measurements of the large-scale ($>\lambda$) surface
topography. This method is fast enough to characterize each individual
fiber in reasonable time.

\begin{figure}[tb]
  \centering
  \includegraphics[width=0.75\textwidth]{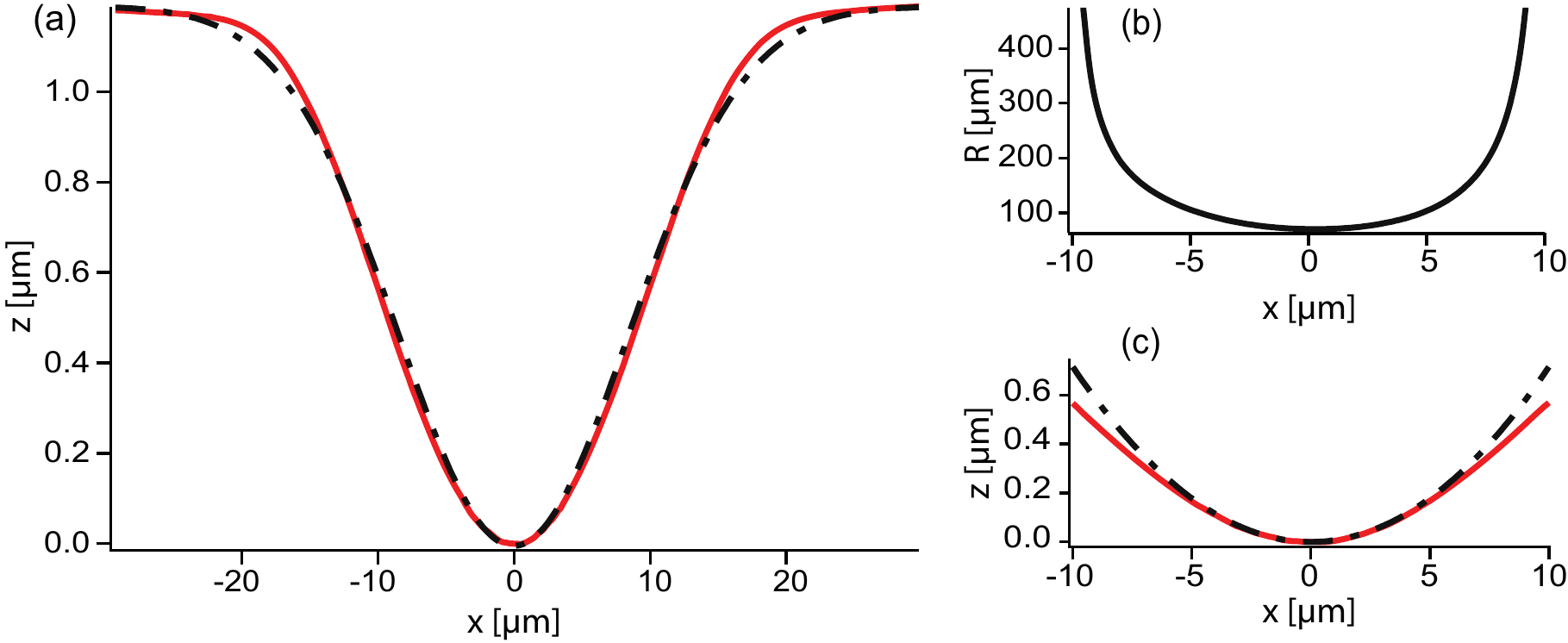}
  \caption{(a) Cut through a profile obtained with the interferometric
  microscope (red solid line), and its fit by a Gaussian (black dot-dashed
  line). (b) Local ROC as a function of the transverse
  coordinate $x$, obtained from a tenth-order polynomial fit to the
  data in (a) to reduce the effect of measurement noise. (c) Zoom of
  the data in (a) (red solid line) and a section of a circle with
  $R=70\,\mu$m (the ROC of the profile at $x=0$) for comparison (black
  dot-dashed line).}
  \label{fig:profile}
\end{figure}

\begin{figure}[htb]
  \centering
  \includegraphics[width=0.25 \textwidth]{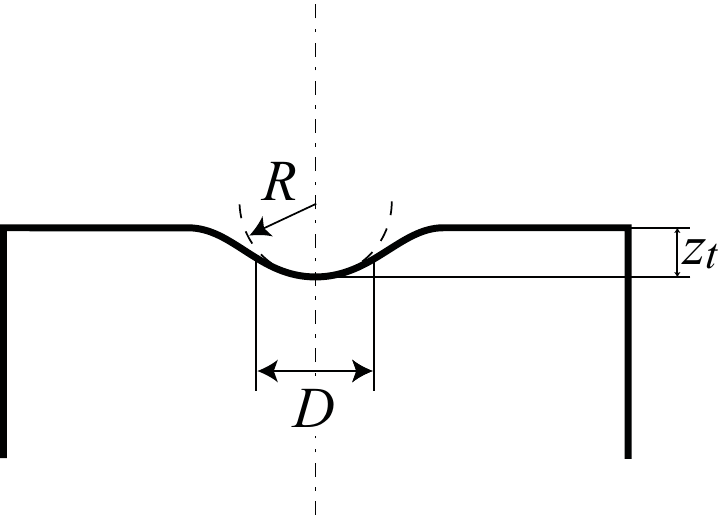}
  \caption{Mirror geometry and parameters. The profile is 
           not spherical; $R$ designates the ROC in the 
           center and $D$ the structure diameter as defined in the
           text. $z_t$ is the depth of the structure.
    \label{fig:RDDef}} 
\end{figure}

Figure~\ref{fig:surface} shows SEM and AFM images of representative
fiber end faces after the application of a single laser pulse.  The
section through the center in fig.~\ref{fig:profile}(a) shows a
profile which is reasonably well approximated by a gaussian,
$z(\rho)=z_t \exp(-\rho^2/\sigma^2)$ over a wide parameter range. (A
limit is reached for long pulses and big waists, where a rim begins to
form around the depression.) Because the profile is not spherical, the
local ROC varies with the transverse coordinate
(fig.~\ref{fig:profile}(b)). Close to the center of the profile, this
variation is slow however, and the shape is well approximated by a
circle (fig.~\ref{fig:profile}(c)). We use the central ROC
to define the mirror curvature $R$ (fig.~\ref{fig:RDDef}).  The full
width at $1/e$ of the gaussian profile gives an estimate of the
useful mirror diameter $D$ (cf. \ref{sec:Finesse}).

Because of the approximately gaussian shape of the depression, $R$,
$D$ and the total structure depth $z_t$ are related by
\begin{equation}
  z_t\approx\frac{D^2}{8 R}\,.
\end{equation} 
For example, a mirror structure with $R=50\,\mu$m and useful diameter
$D=10\,\mu$m is only $z_t=0.25\,\mu$m deep.  With the laser parameters
given above, resulting structures are 0.01 to 4\,$\mu$m deep and have
diameters $D$ between 10 and $45\,\mu$m. ROCs measured
at the bottom of the depression are between $40$ and
$2000\,\mu$m. Part of the processed surfaces show a
slight ellipticity (up to a few percent), which is probably
due to an observed astigmatism of the CO$_2$ beam: when varying the
fiber position along the CO$_2$ beam axis, the beam profile changes
from circular to elliptical. Alignment error along this axis
translates into ellipticity of the depression. This could easily
be improved in a future setup. 

AFM measurement areas were between $(500\,\textrm{nm})^2$ and
$(10\,\mu\textrm{m})^2$. To extract surface roughness, we subtract a
two-dimensional, fourth-order polynomial that accounts for the concave
overall shape and then calculate the isotropic power spectral density
(PSD). Integrating the PSD up to $2/\lambda$ consistently yields
$\sigma_{\textrm{\small{sc}}}\sim 0.2\pm0.01\,$nm. (Reference
measurements on mica sheets (neglegible roughness) yield
$\sigma_{\textrm{noise}}=0.1\,$nm.)  A widely used estimate linking
this roughness to the scatter loss is \cite{Bennett92}
\begin{equation}
\S\approx\left( \frac{4\pi\sigma_{\textrm{\small{sc}}}}{\lambda}\right)^2\,.
\label{eq:scattloss}
\end{equation}
We thus obtain an estimate of $\S=10\,$ppm for near-infrared light at
$\lambda=780\,$nm, assuming a high-quality mirror coating that does
not significantly increase this roughness. Such a coating also has very
low absorption loss, $\A=2\,$ppm being a realistic value
\cite{Hood01}. With a transmission equaling the losses, $\T=\S+\A$, we
thus expect a maximum finesse $\Fi=\pi/(\T+\S+\A)\approx 130000$ for a
cavity made from two fiber mirrors with identical coatings, and 170000
if the transmission of one mirror is reduced to 2\,ppm.

\section{Cavity and coupling parameters of FFPCs}

\subsection{Definitions and optical parameters}

We consider cavities consisting of two mirrors, labeled $i=1,2$, of
intensity transmission $\T_i$, scatter and absorption losses
$\L_i=\S_i+\A_i$, and reflectivity $\R_i=1-(\T_i+\L_i)$. ROCs are
$R_i$ and (effective) diameters $D_i$. The optical distance between
the mirrors is $L$ (which is slightly larger than the geometric
distance $L_{\textrm{geom}}$, cf.~sec.~\ref{sec:waist}). Basic
quantities characterizing the cavity are its free spectral range
$\FSR=2\pi c/(2L)$ and
the width of the TEM$_{00}$ cavity resonances, usually expressed as FWHM
frequency $\delta\nu$ in laser physics and as HWHM angular frequency
\begin{equation}
\kappa=\frac{2\pi\delta\nu}{2}=\frac{c\LTot}{4L}=\frac{\pi c}{2L\Fi}
\label{eq:kappa}
\end{equation}
in CQED. Here, $\LTot$ is the round-trip loss
($\LTot=\T_1+\T_2+\L_1+\L_2$ if there are no additional losses such as
clipping). The cavity finesse is
\begin{equation}
\Fi=\frac{\FSR}{2\kappa}=\frac{2\pi}{\LTot}\,.
\label{eq:Fi}
\end{equation}
In contrast to the quality factor $Q=\nu/\delta\nu=\Fi L/(\lambda/2)$,
the finesse depends only on the properties of the mirror coatings and
not on the cavity length (as long as the mirror diameters $D_i$ are
large enough to neglect clipping loss -- see below).

\subsection{Waist radius}
\label{sec:waist}

\begin{figure}[tb]
  \centering
  \includegraphics[width=0.9\textwidth]{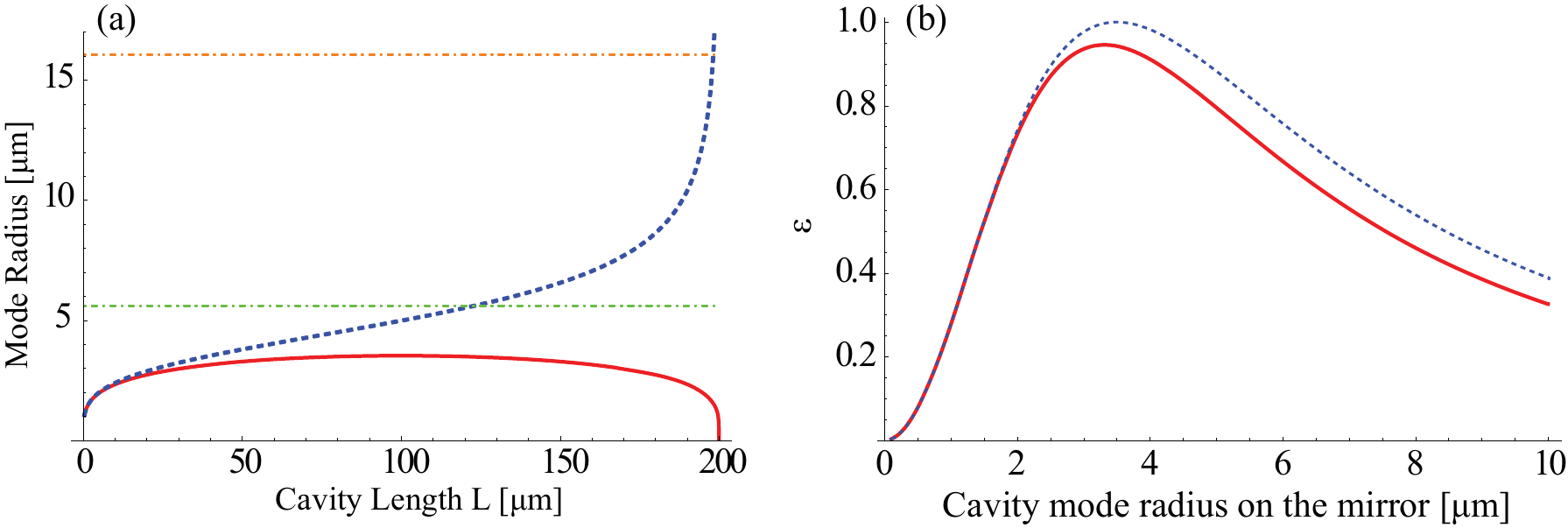}
  \caption{(a) Waist radius $w_0$ (red solid line) and mode
    radius $w_{1,2}$ on the mirrors (blue dashed line) for a cavity
    with $R_1=R_2=100\,\mu$m and $\lambda=780\,$nm. Dot-dashed lines
    indicate the maximum $w_{1,2}$ for a mode-matching efficiency
    $\mmeff=0.8$ to a fiber with mode field radius $w_f=3.5\,\mu$m
    (green line) and for neglegible clipping loss up to $\Fi=150000$
    assuming mirror diameter $D=45\,\mu$m (orange line).
    (b) Power coupling efficiency $\mmeff$ between a SM
    fiber (mode field radius $w_f=3.5\,\mu$m) with a mirror of
    curvature $R=150\,\mu$m and the cavity mode, as a function of
    the cavity mode radius $w_m$ on the mirror. Red, solid line: Full
    model taking into account wavefront curvature and the lensing
    effect of the mirror surface (eq.~(\ref{eq:mmFull})). Blue, dashed
    line: wavefront curvature and the lensing effect are neglected
    (\ref{eq:mmeff})). The approximation is very good for small $w_m$,
    and remains remarkably accurate even for larger values. The
    predicted maximum efficiency is $\mmeff=1$ for $w_m=3.5\,\mu$m in
    the approximation and $\mmeff\approx 0.95$ for $w_m\approx
    3.3\,\mu$m in the full expression.
    \label{fig:fceff}}
\end{figure}

In many applications, the foremost requirement is a small waist
$w_0$ in order to optimize coupling to an emitter located inside the
cavity. In the symmetric case $R_1=R_2=R$,
\begin{equation} 
  w_0=\sqrt{
    \frac{\lambda}{2\pi} } \left( L \left(2R - L \right)
    \right)^{\frac{1}{4}}\,.
\label{eq:waist}
\end{equation}
(fig.~\ref{fig:fceff}, left). With macroscopic supermirrors, the
interesting region $w_0\lesssim 5\,\mu$m is only accessible in the
near-concentric regime $L\sim 2 R$, which is difficult or impossible
to exploit due to the extreme alignment sensitivity in this
regime. Existing macroscopic FP cavities (FPCs) rather have $w_0\sim
20\,\mu$m.  The FFPC design makes it possible to enter the interesting
regime of small $L$ and small $R$ simultaneously, which is
inaccessible to macroscopic cavities.  Typical $R$ values for our
laser-machined fiber mirrors are in the $100\,\mu$m range, two to
three orders of magnitude smaller than for traditional high-finesse
FPCs.  Additionally, for a given $R$, the length of the fiber cavity
can be made much smaller than for its macroscopic counterpart, because
of the smaller mirror diameter. (The length limit is reached when the
mirrors touch \cite{Hood01}.)  Both factors contribute to enable
exceptionally small waists for an open cavity.

The full expression of $w_0$ for non-symmetric cavities can be found
in textbooks such as \cite{Siegman86}. As with symmetric cavities,
small waists are obtained for short cavities ($L\ll R_1,R_2$) and
close to the stability limits ($L\approx R_1+R_2$ and $L\approx
|R_1-R_2|$).  For short cavities, the full expression simplifies to
\begin{equation}
  w_0\approx \sqrt{\frac{\lambda}{\pi}}\left(L \frac{R_1 R_2}{R_1+R_2}\right)^{\frac{1}{4}}\,.
\end{equation}
If furthermore $R_1$ and $R_2$ are substantially different, the
minimum waist size is determined by the smaller of the two. For example,
if $R_1\ll R_2$, then
\begin{equation}
w_0\approx \sqrt{\frac{\lambda}{\pi}}\left(L R_1\right)^{\frac{1}{4}}  \,.
\end{equation}
This limit particularly applies to half-symmetric cavities
($R_2=\infty$). Interestingly, comparison with eq.~\ref{eq:waist}
shows that when replacing one mirror of a short symmetric cavity by a
planar one (leaving $L$ unchanged) increases $w_0$ by less than
$20\,\%$.

To summarize: if we exclude cavities close to the
stability limit, then a small-waist cavity should be as short as
possible and have at least one strongly curved mirror.

\subsection{Minimum length and waist}
\label{sec:minlength}
To determine the smallest possible $w_0$ within our fabrication
limits, we have to consider how small $L$ can be made. We can
currently machine mirror structures with diameters down to $10\,\mu$m.
As we have seen in sec.~\ref{sec:surface}, the mirror depth is about
$z_t=250\,$nm for an $R=50\,\mu$m fiber mirror with $D=10\,\mu$m, so
that a geometric cavity length as short as $L_{\textrm{geom}}=500\,$nm
can be realized even with such a strongly curved mirror. (Note that
the $D=10\,\mu$m is still large enough to avoid clipping loss,
cf. sec.~\ref{sec:clip}.) The part of the field penetrating into the
multilayer stack then contributes significantly to the effective
cavity length. We account for this effect by setting
$L=L_{\textrm{geom}}+\alpha_{\textrm{ml}}\lambda/2$ \cite{Hood01},
where the coating materials give $\alpha_{\textrm{ml}}\simeq 1.6$ in
our case -- dominating over the minimum geometric length. Taking into
account this effect and leaving a small gap to introduce atoms, it
should still be possible to achieve $L\approx 2\mu$m for all $R$ down
to our current fabrication limit $R=50\,\mu$m. A symmetric cavity with
$R=50\,\mu$m and $L=5\lambda/2\approx 2.0\,\mu$m then has
$w_0=1.3\,\mu$m at $\lambda=780\,$nm according to
eq.~(\ref{eq:waist}). This is more than ten times smaller than the
waist of macroscopic CQED cavities and slightly smaller than that of
high-Q micropillars \cite{Reitzenstein07}. Modeling this mode more
accurately would require taking into account light propagation inside
the multilayer stack (and not just the stack's effect on cavity
length), and may also require going beyond the paraxial approximation
\cite{vanEnk01}.

Minimizing $L$ and $R$ also minimizes the mode volume
\begin{equation}
  V_m=\frac{\pi}{4} w_0^2 L  =\lambda/8 \sqrt{2R L^3 - L^4}\,,
\label{eq:modevol}
\end{equation}
where the second form is for the symmetric case.  
With the parameters above, $V_m\approx 2.6\,\mu\mbox{m}^3\approx 5.5
\lambda^3$. $V_m$ enters in CQED coupling rates, which will be
discussed below in sec.~\ref{sec:CQEDparams}.

\subsection{Fiber coupling}
\label{sec:coupling}

As the cavity mirrors are part of the incoupling and outcoupling
fibers, coupling to and from the cavity is robust and stable over
time. This is one key advantage of the FFPC.  There are no
mode-matching lenses, so coupling efficiency is given directly by the
mode matching between the mode leaving the fiber and the cavity
mode. If single-mode operation is not required -- such as for the
output fiber in a cavity transmission measurement -- a multimode (MM)
fiber can be used on the output side \cite{Colombe07}. This virtually
eliminates coupling loss and also makes the cavity more robust against
various types of misalignment, such as centering errors between the
mirror and the fiber core. But even for SM fibers, the coupling
efficiency $\mmeff$ is typically very high, in spite of the strong
mirror curvature. For example, we have measured $\mmeff>85\%$ for a
SM fiber and a mirror with $R=450\,\mu$m (see
sec.~\ref{sec:couplingExp}).

This high efficiency can be seen as a consequence of the small cavity
mode radius, which leads to small phase difference across the relevant
mode cross-sections even for strongly curved wavefronts. To get a
feeling for the orders of magnitude, consider a spherical wavefront
with ROC $R\gg\lambda$. This wavefront deviates from
its tangential plane by $\lambda/2$ when the transverse coordinate
$\rho$ reaches $\rho_{\lambda/2}\lambda\approx\sqrt{R/\lambda}$, so
even on an $R=100\,\mu$m mirror and $\lambda=1\,\mu$m, the mode radius
can be as large as $10\,\mu$m before it deviates from planar by
$\lambda/2$. Typical mode field radii in SM fibers are much smaller
than this. This has several simplifying consequences: First, the
lensing effect by the curved fiber surface can be neglected, and
second, the power coupling efficiency $\mmeff$ between the fiber and
cavity modes can be approximated simply by the overlap integral of the
fiber and cavity mode intensity distributions, neglecting phase
mismatch. For a SM fiber with its nearly gaussian transverse mode
profile of radius $w_{f}$, and assuming that the cavity mode is well
aligned with the fiber axis, we thus obtain the simple estimate
\begin{equation}
\mmeff\approx\left(\frac{2w_f w_m}{w_f^2+w_m^2}\right)^2\,,
\label{eq:mmeff}
\end{equation}
where $w_m$ is the cavity mode's radius on the mirror. A more complete
expression including lensing effect and wavefront curvature is 
\begin{equation}
 \mmeff=
    \frac{4}
      {\left(\frac{w_f}{w_m}+\frac{w_m}{w_f}\right)^2+
            \left(\frac{\pi n_f w_f w_m}{\lambda R}\right)^2
      }\,,
\label{eq:mmFull}
\end{equation}
where $n_f$ is the refractive index of the fiber and $R$ the ROC of the mirror. This expression is derived in the appendix. What it still ignores is misalignment between the mirror and the fiber axis.

In figure \ref{fig:fceff}(right), $\mmeff$ is plotted as a function of
$w_m$. Optimum coupling occurs around $w_m= w_f$. The remaining
mismatch is due to mirror curvature. To minimize it, $R$ should be
chosen as large as possible. $\mmeff=1$ is achieved on the planar side
($R=\infty$) of a half-symmetric\footnote{Note that the geometric
  cavity length must be slightly shortened due to
  finite coating thickness.}  cavity 
where $L$ and the curvature of the second mirror are chosen such that
$w_m=w_{f}$. A possible drawback of this configuration is its waist
size, $w_0=w_f$: $w_f$ is typically much larger than the minimum $w_0$
which FFPCs can achieve.

If a MM fiber is used on the outcoupling side, the coupling
coefficient is determined by its numerical aperture $NA=\sin
\Theta_{\textrm{acc}}$. $\Theta_{\textrm{acc}}$ is the acceptance
angle of the fiber, which must be compared to the full divergence
angle of the cavity mode. The latter grows from 0 (in the waist) up to
$2\lambda/(\pi w_0)$ (for $z\gg z_R$). As an example, for a typical
$NA=0.1$ at $\lambda=780\,$nm, in the worst case of a long cavity
($L\gg z_R$), the waist can still be made as small as $w_0=2.4\,\mu$m
before the mode divergence becomes larger than the acceptance angle of
the fiber. The large acceptance angle also makes the cavity more robust with respect to misalignment.

\subsection{Maximum length and clipping loss}
\label{sec:clip}

In some applications, a larger mirror distance is required, for
example to introduce an object such as a membrane \cite{Thompson08}),
to minimize trap distortion and heating in an ion trap, or to reduce
the FSR.  As $L$ is increased, the mode radius on at least one of the
mirrors will also increase, and this limits the cavity length through
two effects: first, fiber coupling efficiency decreases, which may or
may not be a problem depending on the application. Then, cavity
finesse decreases due to clipping loss.

In this section, we will consider these effects for just one of the fiber
mirrors, calling $D$ the mirror diameter, $w_m$ the cavity mode radius
on this mirror, and $w_{f}$ the mode field radius in its fiber. We will first determine the maximum allowable $w_m$ as imposed by (a)
fiber coupling and (b) clipping. The maximum $L$ for a given $w_m$ then
follows from the standard FPC formulas (c).

(a) Calling $\mmeffmin$ the target fiber coupling efficiency, and
using the approximation of eq.~(\ref{eq:mmeff}), $w_m$ must fulfill
\begin{equation}
  w_m\le w_{\mmeff}:= 
  w_{f}\left(\frac{1}{\sqrt{\mmeffmin}}+\sqrt{\frac{1-\mmeffmin}{\mmeffmin}}\right)   \,.
\end{equation}
For example, if the fiber coupling efficiency is to be at least
$\mmeffmin=0.8$, the cavity mode radius must not
be larger than $1.6\,w_{f}$. For the SM fibers used in our
cavities ($w_f=3.5\,\mu$m), this estimate yields $w_{\mmeff}=5.6\,\mu$m.

(b) To estimate clipping loss on the the finite-diameter fiber mirrors, we
conservatively\footnote{The actual cavity mode for finite-diameter
  mirrors tends to have less power in the periphery than the gaussian
  mode \cite{Siegman86}.} assume a gaussian cavity mode and consider
its ``spillover'' loss upon reflection on a finite-diameter
mirror. For a single reflection,
\begin{equation}
\LCl=e^{-2(D/2)^2/w_m^2}\,,
\label{eq:cliploss}
\end{equation}
where $w_m$ is the mode radius of the mode impinging on the mirror of
diameter $D$.
 If $\LCl$ is to contribute less than 10\% to the total
loss,
\begin{equation}
\LCl < \frac{\LTot}{10}=\frac{2\pi}{10\Fi}\,,
\end{equation}
the mode radius on the mirror must verify
\begin{equation}
  w_m\le w_{\textrm{cl}}:=\clipfac \frac{D}{2} \qquad\textrm{with}\qquad
  \clipfac=\sqrt{\frac{2}{-\ln\left(\frac{2\pi}{10\Fi}\right)}}\,,
\end{equation}
In the finesse range of interest here, $\clipfac^{-1}\approx 2.2 -
2.5$.  With our current fabrication limit of $D\le 45\,\mu$m, we thus
have $w_{\textrm{cl}}\approx 18\,\mu$m in the most demanding case
($\Fi\approx 130000$). This limitation is generally less restrictive
than the one imposed by good fiber coupling.

(c) Now let us calculate the maximum $L$ by requiring that the actual
$w_m$ be smaller than these maximum values. From the elementary FPC
formulas, we have for a symmetric cavity
\begin{equation}
w_m^2=w_0^2\left(1+\left(\frac{\lambda L}{2\pi
      w_0^2}\right)^2\right)
       =\frac{L\lambda}{\pi}
         \sqrt{\frac{1}{1-\left(1-\frac{L}{R}\right)^2}}\,.
\label{eq:w12sym}
\end{equation}
For a given pair of identical mirrors, the maximum length is
\begin{equation}
L_{\max}=2R \frac{1}{1+\left(\frac{\lambda R}{\pi w_{\max}^2}\right)^2},
\end{equation}
with $w_{\max}=w_{\textrm{cl}}$ or $w_\mmeff$ depending on the
case. Obviously, if $w_{\max}^2\lesssim \lambda R$, this becomes much
shorter than the stability limit.

If the application also imposes $w_0$, the length limit follows from
the first expression in eq.~(\ref{eq:w12sym}):
\begin{equation}
  L_{\max}=\frac{2\pi}{\lambda}w_0 \sqrt{w_{\max}^2-w_0^2}
\,.
\end{equation}
For example, with $w_{\max}=16\,\mu$m (low clipping loss) and
$\lambda=780\,$nm, we find that a cavity with $w_0=2\,\mu$m can be
made up to $L_{\max}=125\,\mu$m long. The curvature of the cavity
mirrors is then $R=67\,\mu$m. Requiring instead $w_{\max}=5.6\,\mu$m
(good mode matching) leads to $L_{\max}=84\,\mu$m and $R=48\,\mu$m.

If there are no restrictions on the mode waist, as might be the case
for a filter cavity or a membrane cooling experiment for example, $R$
can be optimized to further increase $L$ while keeping $w_{1,2}$ small
(right expression in eq.~(\ref{eq:w12sym})). The length limit for the
symmetric cavity case then becomes
\begin{equation}
  L_{\max}=\frac{\pi}{\lambda}w_{\max}^2
\end{equation}
and is attained for a confocal cavity ($R=L$). In other words, for a
given, long $L$, the confocal configuration has the smallest mode
radii on the mirrors, and therefore optimizes both mode matching and
finesse.
Taking again $\lambda=780\,$nm and $w_{\max}=16\,\mu$m (low clipping
loss), we find $L_{\max}=1\,$mm, which is of the same order as the
limit imposed by the stability criterion for our maximum ROC. Thus,
within this simple model, clipping loss in confocal cavities remains
neglegible except for very long wavelength and the confocal cavity
length is limited to $L\lesssim 500\ldots 1000\,\mu$m by the
attainable ROC. Note however that our simple model does
not include other imperfections, such as misalignment between the
fiber axis and the center of the mirror. In any case, it should be
possible to make still longer cavities from one fiber mirror and one
macroscopic mirror with larger diameter and ROC.

The more restrictive requirement $w_{\max}=5.6\,\mu$m (good mode
matching) leads to $L_{\max}=126\,\mu$m. This is the longest cavity
that can be made if the power transmission between fiber and cavity
modes is to be at least $\mmeffmin=0.8$. For comparison, the $L=500\,\mu$m
confocal cavity has $\mmeffmin=0.32$, which is still an acceptable
value in many cases.
For $R_1\ne R_2$, $w_i$ still generally grows with $L$ except in the
regions near $L=R_1$ and $L=R_2$, where the resonator becomes
unstable.

\subsection{CQED parameters as functions of the cavity parameters}
\label{sec:CQEDparams}

In cavity QED problems, the cavity parameters enter in the form of the
coherent single-photon coupling rate $g$ and the incoherent cavity
decay rate $\kappa$. It is instructive to consider these rates as
functions of $w_0$ and $L$ (choice of a gaussian mode), and alternatively
as functions of $R$ and $L$ (choice of a pair of mirrors), where we restrict
ourselves to the symmetric case for simplicity. 

\begin{figure}[tb]
  \centering
    \includegraphics[width=0.45\textwidth]{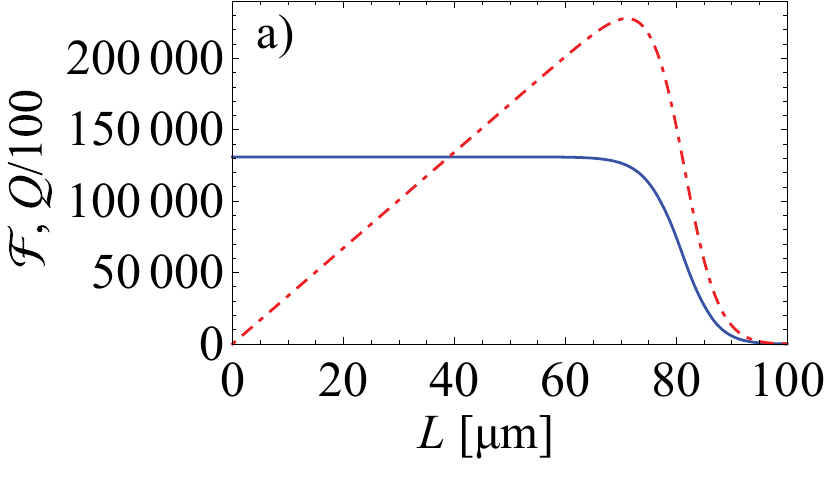}
    \includegraphics[width=0.45\textwidth]{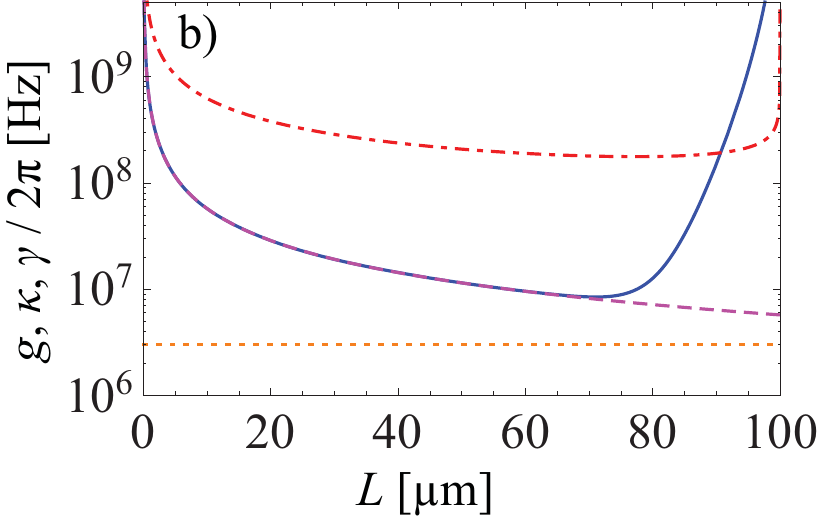}
    \includegraphics[width=0.45\textwidth]{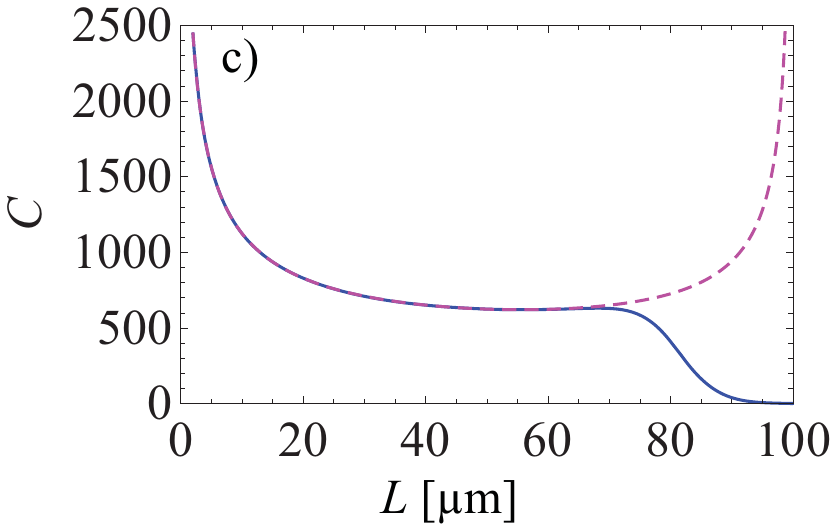}
    \includegraphics[width=0.45\textwidth]{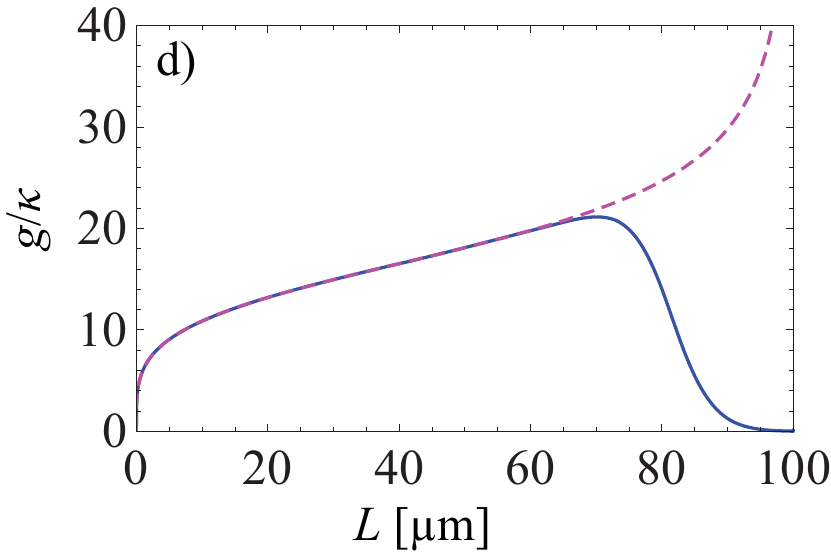}
  \caption{CQED parameters taking into account clipping loss on
    finite-diameter mirrors, for a cavity with the following parameters:
    $R_1=100\,\mu\mbox{m},R_2=450\,\mu\mbox{m}, D_1=D_2=30\,\mu\mbox{m}$,
    coupling to Rb atoms 
    ($\lambda=780\,\mbox{nm},\gamma/(2\pi)=3\,\mbox{MHz}$). (a)
    Finesse (blue solid line) and Q factor (red dot-dashed
    line). The decay for
    $L\gtrsim 75\,\mu$m is caused by clipping, and is responsible for
    the increase in $\kappa$ and decrease in $C$ at these lengths. (b)
    $g_0$ without clipping (red, dot-dashed), $\kappa$ without clipping
    (magenta, dashed) and with clipping loss (solid,
    blue). Throughout
    the length range, $\kappa$ is larger than $\gamma$ (orange,
    dotted). (c) $C$ without clipping (magenta, dashed) and with
    clipping (blue, solid). Even without clipping loss, the high
    cooperativity near $L=100\,\mu$m would be difficult to reach in
    practice, as it occurs at the limit of the stability range. (d)
    The ratio $g_0/\kappa$ without (magenta, dashed) and with clipping
    (blue, solid). As $\kappa>\gamma$, this ratio quantifies the
    strong coupling. In contrast to $C_0$, it increases for longer
    cavities as long as clipping is neglegible.
    \label{fig:CQEDParams}}
\end{figure}

For a cavity mode with frequency $\omega$ and an atom or other emitter
(dipole matrix element $\mu$) located at maximum field intensity, the
coherent coupling rate is
\begin{equation}  
  g_0=\sqrt{\frac{\mu^2\omega}{2\hbar\epsilon_M V_m}}
     =\sqrt{\frac{3\lambda^2 c\gamma}{4\pi V_m}}\,,
\end{equation}
where $\gamma$ is the HWHM linewidth of the excited state,
$\epsilon_M$ is the dielectric constant at the location of the emitter
($\epsilon_M=\epsilon_0$ for free space). The second form applies to a
two-level atom in free space; for simplicity we will use this form
from now on. $g_0$ depends on the cavity parameters through the mode
volume ((eq.~\ref{eq:modevol})). Fig.~\ref{fig:CQEDParams} shows an example.

We have already determined the minimum achievable mode volume,
$V_m\approx 5.5 \lambda^3$ for $\lambda=780\,$nm
(sec.~\ref{sec:minlength}). This leads to a maximum coupling rate of
$g_0=2\pi\times 2.8\,$GHz for Rb atoms, about four times the most
optimistic projected limit \cite{Hood01} of macroscopic FPCs. Note
that the limit of \cite{Hood01} assumes significant future
improvements in polishing technology, while our value is calculated
with the surface quality and ROC that have already been fabricated and
measured.

Maximizing $g_0$ requires minimizing $V_m$, so reducing the mirror
spacing (and adapting the ROC accordingly) maximizes $g_0$ for a given
$w_0$. However, doing so also increases the cavity decay rate
(eq.~\ref{eq:kappa}).  If the goal is to enter as far as possible into
the ``strong coupling regime'' (i.e., $g_0>\kappa,\gamma$, or
equivalently, resolved coupled-system resonances), the
coupling-to-dissipation ratio $g_0/\max (\gamma,\kappa)$ may be used
as a figure of merit. If the cavity is short -- more precisely, if
$L<\pi c / (2\Fi\gamma)$ --, then $\kappa>\gamma$, so that this ratio
is optimized by maximizing $g_0/\kappa$. With $\Fi=10^5$ and
$\gamma/2\pi=3\,$MHz, $\pi c / (2\Fi\gamma)=250\,\mu$m, which means
that typically $\kappa>\gamma$ over the full stability
range. Expressing this ratio as a function of the mode and mirror
parameters:
\begin{equation} 
  \frac{g_0}{\kappa}
  =\frac{2\lambda}{\pi^2}\sqrt{\frac{3\gamma}{c}}\,\frac{\Fi\sqrt{L}}{w_0}
  =\frac{2\sqrt{6}}{\pi^{3/2}}\sqrt{\frac{\gamma\lambda}{c}}\frac{\Fi\sqrt{L}}{\left(L(2R-L)\right)^{1/4}}\,.
\end{equation} 
Fig.~\ref{fig:CQEDParams} shows an example. As expected, choosing a small $w_0$ improves this ratio.  Somewhat surprisingly, the ratio also improves with growing cavity length, despite the growing mode volume. This is because $\kappa$ depends on $L$ more strongly than does $g_0$. For long cavities, the ratio is determined by $g_0/\gamma$, and decreases again. 

By contrast, in a large class of applications which notably contains
single-photon sources, a more important figure of merit is the
single-atom cooperativity $C_0$,
\begin{equation}
C_0=\frac{g_0^2}{2\kappa\gamma}=\frac{3\lambda^2}{\pi^3}\frac{\Fi}{w_0^2}
   =\frac{6\lambda}{\pi} \frac{\Fi}{\sqrt{2R L - L^2}} \,.
\label{eq:C0_2lvl}
\end{equation}
$C_0$ determines the Purcell factor; the probability for a spontaneous
photon to be emitted into the cavity mode is $C_0/(C_0+1)$. The inverse of
$C_0$ is called the critical atom number.

In contrast to $g_0$, $C_0$ does not depend on the mode volume, but
only on the mode waist $w_0$. Another way of seeing this is by
realizing that $C_0$ is effectively the optical density of the
emitter, which depends on the ratio of its absorption cross-section to
the cross-section of the mode, but is independent of the cavity length
(as long as $L$ is larger than the emitter size). The conclusion is
that the choice of a particular gaussian mode fixes the value of
$C_0$, no matter where we decide to place the mirrors that confine
this mode. The strong curvature of our fiber mirrors enables very small
waist size, we have already seen that $w_0=1.3\,\mu$m is a realistic
value.
Combined with the maximum finesse $\Fi\approx 130000$
(sec.~\ref{sec:surface}), eq.~(\ref{eq:C0_2lvl}) predicts $C_0=4560$
for the maximum cooperativity that can be achieved at
$\lambda=780\,$nm.

To conclude this section, fig.~\ref{fig:CQEDParams} shows CQED
parameters as a function of cavity length $L$, taking into account
clipping loss, for the following parameters:
$\lambda=780\,\mbox{nm},\gamma/(2\pi)=3\,\mbox{MHz},
R_1=100\,\mu\mbox{m},R_2=450\,\mu\mbox{m}, D_1=D_2=30\,\mu\mbox{m}$
and $\L=\T=12\,$ppm per mirror.

\section{Experimental results}

We have measured transmission and reflection spectra as well as
insertion loss for a variety of FFPCs with mirrors of various
curvatures fabricated on SM and MM fibers. All of them
had coatings fabricated by Laserzentrum Hannover (LZH)
\footnote{Laserzentrum Hannover e.V., Abt. Laserkomponenten, D-30419
  Hannover, Germany}, designed for maximum reflectivity at
780\,nm. Note that these are not the best commercially available
coatings. We have chosen this supplier as a compromise between coating
quality and delay at the time we needed the coatings; the measurements
below indicate that still better cavity performance can be achieved by
coating the same fibers with ``supermirror'' coatings, as in
\cite{Hood01} for example.  

The most comprehensive measurements were done on two cavities, FFP1
and FFP2, which are part of an atom chip experiment in Paris in which
we investigate CQED effects with $^{87}$Rb Bose-Einstein condensates
\cite{Colombe07} and detection of trapped single atoms. These cavities
have been in vacuum for more than 2 years now; periodic remeasurements
show no degradation of their performance in spite of their exposition
to low-pressure Rb vapor.

While measuring the transmission and reflection spectra is fairly
straightforward, other measurments such as insertion loss and mirror
losses tend to be more difficult than with macroscopic cavities due to
the inherent fiber coupling. In the following subsections we describe
our methods and show results for the various parameters.

\subsection{Fiber preparation and coating}
\label{sec:fiberPrep}
We have laser-machined a batch of SM\footnote{Oxford
  Electronics SM800-125CB} and MM gradient index
fibers\footnote{Oxford Electronics GI50-125CB} with $125\,\mu$m
cladding diameter and $7\,\mu$m mode field diameter / $50\,\mu$m core diameter, respectively. The fibers
are metal coated (Cu / Cu alloy), which makes them suitable for
ultra-high vacuum use and also shields stray light. Structures with
ROC from $90\,\mu$m to $500\,\mu$m and mirror
diameters from $20\,\mu$m to $40\,\mu$m were coated in an ion beam
sputtering process by LZH. Each fiber was cleaned in aquaeus solution
of HCl (H$_2$O:HCl 5:1) for 2 min in an ultrasonic bath (USB), then
rinsed for 2 min (USB) in H$_2$O and finally for 2 min (USB) in
acetone. After cleaning, the surfaces were individually inspected for
contaminations and then inserted into a purpose-built holding plate. This
procedure was carried out immediately before coating, but outside the
cleanroom in which the coating took place.

The dielectric coating has 14 layers of SiO$_2$ with refractive index
$n_{SiO}=1.455$ and 15 layers Ta$_2$O$_5$ with $n_{TaO}=2.105$. The
calculated transmission of the layer stack is $\T=34\,$ppm at
$780\,$nm. Reference substrates with $\sigma\sim0.1\,$nm rms roughness
were coated in the same run.

A microscope inspection of several coated fibers showed that 30-50\% of
the end faces contained contaminations, some of them making the fibers
unemployable. The most likely explanation is that dust remaining on
the holding plate contaminated the fibers when they were inserted.

\subsection{Coating transmission and losses}
\label{sec:coatingProps}
To determine the quality of the coatings, we have done various
measurements both with reference substrates and with fibers.  The LZH
carried out a calorimetric absorption loss measurement at 1064nm which
yielded $\A=100\pm30\,$ppm and an Ulbricht sphere total scatter
measurement at 633\,nm which gave $\S=75\pm20\,$ppm. Our AFM
measurements of the coated reference substrates show a roughness of
$\sigma_{sc}=0.20\,$nm rms, an increase of at least 0.1\,nm over the
value before coating and corresponding to scatter loss
$\S=10\,$ppm. Our direct measurement of the transmission of reference
substrates yielded $\T=31\pm 5\,$ppm, close to the
calculated value.  We also built macroscopic cavities from the
reference substrates and measured their decay time
constant $\tau=\Fi L/\pi c$ using a fast scan
ringdown method \cite{An95}. The resulting value $\tau=3.35\,\mu$s obtained for $L=100\,$mm corresponds to a
finesse $\Fi=31000$ and total loss per mirror
$\T+\L=\pi/\Fi=101\,$ppm. Table \ref{tab:mirrorLosses}
summarizes these results. The deviations between the different
measurement methods suggest isolated defects, possibly the dust
particles mentioned above.

\begin{table}
\begin{centering}
\begin{tabular}{|l|l|r|r|}
\hline
\strut Quantity&Method&Macroscopic substrates&Fibers\\
\hline
\strut Absorption loss $\A$&Calorimetric @ 1064\,nm&$100\pm30\,$ppm&\\
\strut &Bistability&&$30\pm 6\,$ppm\\
\hline
\strut Scatter loss $\S$&Ulbricht sphere @633\,nm&$75\pm 20\,$ppm&\\
\strut &AFM roughness&10\,ppm&$23\pm 3\,$ppm\\
\hline
\strut Transmission $\T$&Direct&$31\pm 5\,$ppm&
\\
\hline
\strut $\T+\A+\S$&Finesse (Fast ringdown)&101\,ppm&\\
\strut &Finesse (Cavity scan)&&$85\pm 12\,$ppm\\
\hline
\end{tabular}
\end{centering}
\caption{\label{tab:mirrorLosses}
  Summary of mirror loss measurements. Measurements were carried out at
  780\,nm unless otherwise noted. The finesse measurements, being
  carried out with two mirrors from the same coating run, provide a
  fairly accurate value for the total loss per mirror$\T+\A+\S$. The individual
  measurements of $\T$, $\A$ and $\S$ should add up to the same value
  within the error margins. This is the case for the fiber
  mirrors. For the macroscopic
  substrates however, the sum of the individual measurements is larger than
  the total value from the finesse measurement. One possible
  explanation would be low-density defects (dust particles) that
  increased the loss in the calorimetric and/or Ulbricht sphere 
  measurement.}
\end{table}

The total losses for the fiber mirrors were obtained from finesse
measurements with short FFPCs, which are discussed in more detail
below. Measurements on several cavities yield $\Fi=37000 \pm 5000$,
corresponding to total loss $\T+\L=\pi/\Fi=85\pm 12\,$ppm. Using the
value $\T=31\pm 5\,$ppm obtained for the macroscopic substrates, we
deduce $\L=(\pi/\Fi)-\T=54\pm17\,$ppm. An independent estimate of $\S$
comes from the AFM measurements. Like the reference substrates, the
coated fibers exhibit an increased surface roughness, in this case to
$\sigma_{sc}=0.32\,$nm rms. Using eq.~\ref{eq:scattloss}, this
corresponds to $\S=23\,$ppm. Absorption loss can be estimated from
absorption-induced bistability of the cavity (see section
\ref{sec:bistability}); we obtained $\A\sim 30$ppm. The sum of these
individual estimates is $\L=\S+\A=53\,$ppm. (Considering the errors of
the individual measurements, the very good agreement with
$(\pi/\Fi)-\T$ is partly fortuitous.)

These measurements consistently indicate that the losses of the FFPCs
are dominated by the coatings. In particular, finesse values
obtained with macroscopic and fiber mirrors are very similar, in spite
of the significantly lower roughness of the macroscopic substrates
before coating. The AFM measurements reveal that, indeed, the coatings
significantly increased the surface roughness. (For state-of-the art
``supermirror'' coatings, it is known that this does not occur
\cite{Hood01}.) There is no indication that the small ROC of the fiber mirrors would cause any significant reduction of the coating quality as compared to the reference substrates. Using state-of-the art coatings should make it possible to fully exploit the surface quality of the laser-machined fibers.

\subsection{Fiber cavity mounting and alignment}

\begin{figure}[tb]
  \centering
  \includegraphics[width=0.49\textwidth]{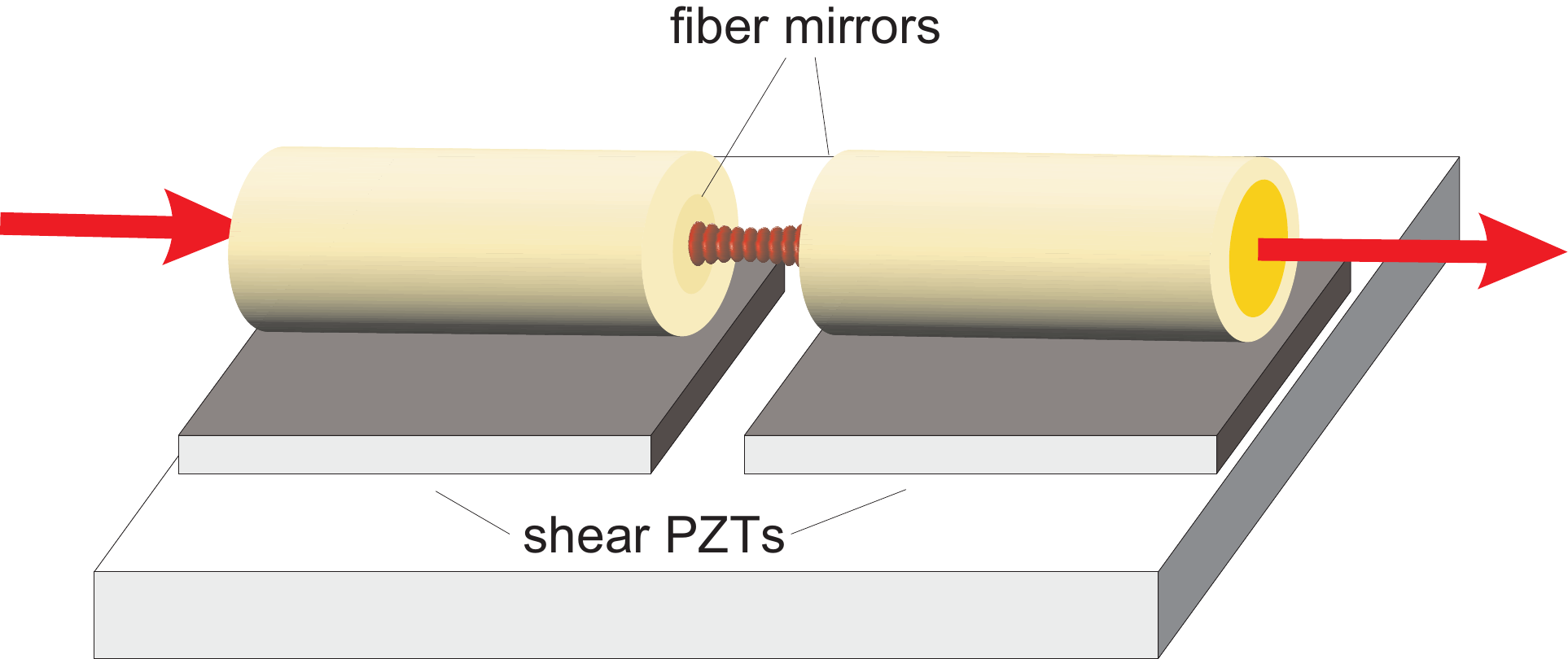}
  \caption{Miniature mount for tunable FFPCs. Piezoelectric transducers (PZTs) are used to scan the cavity length. All other alignment is done during assembly.
  \label{fig:mount}}
\end{figure}

Fibers were angle-cleaved at their coupling ports and mounted either
in a macroscopic mount based on translation stages (``test mount''),
or in a dedicated ``miniature mount'' involving v-grooves and shear
piezos on a ceramic baseplate made from Macor
(figure~\ref{fig:mount}). The miniature mount has better passive
stability, as expected from its low profile and small overall size.

In the test mount, fibers are clamped into side-loaded ferrules that
are fixed on two micropositioning stages. Together they provide three
axes of translation, including fine tunability by integrated piezos,
and two angular degrees of freedom.  For the miniature mount, we use
a similar micropositioning setup to align the fibers during assembly.
After glueing, the only adjustable parameter is cavity length
within the displacement range of the shear piezos, which is several
FSRs. All other alignment is done during assembly, where we apply and
cure epoxy glues while monitoring cavity transmission (``active
alignment'').  We use Epo-Tek 353ND and Epo-Tek 301 to glue the
components together. Both are UHV compatible down to $p <
10^{-10}\,$mbar. Cavity alignment is observed visually with a stereo
microscope (magnification up to 63x) and magnifying video cameras from
two axes. After a geometrical preadjustment, the alignment is
optimized by maximizing the overall transmission of the fiber cavity,
which is scanned over at least one FSR continuously. We did not
observe any degradation of the alignment of the glued cavity over
time, nor did baking to $100^{\circ}$C leave a measurable irreversible
effect once the system had cooled down.

\subsection{Tranmission spectrum, finesse and clipping}

\label{sec:Finesse}
\begin{figure}[tb]
  \centering
  \includegraphics[width=0.49\textwidth]{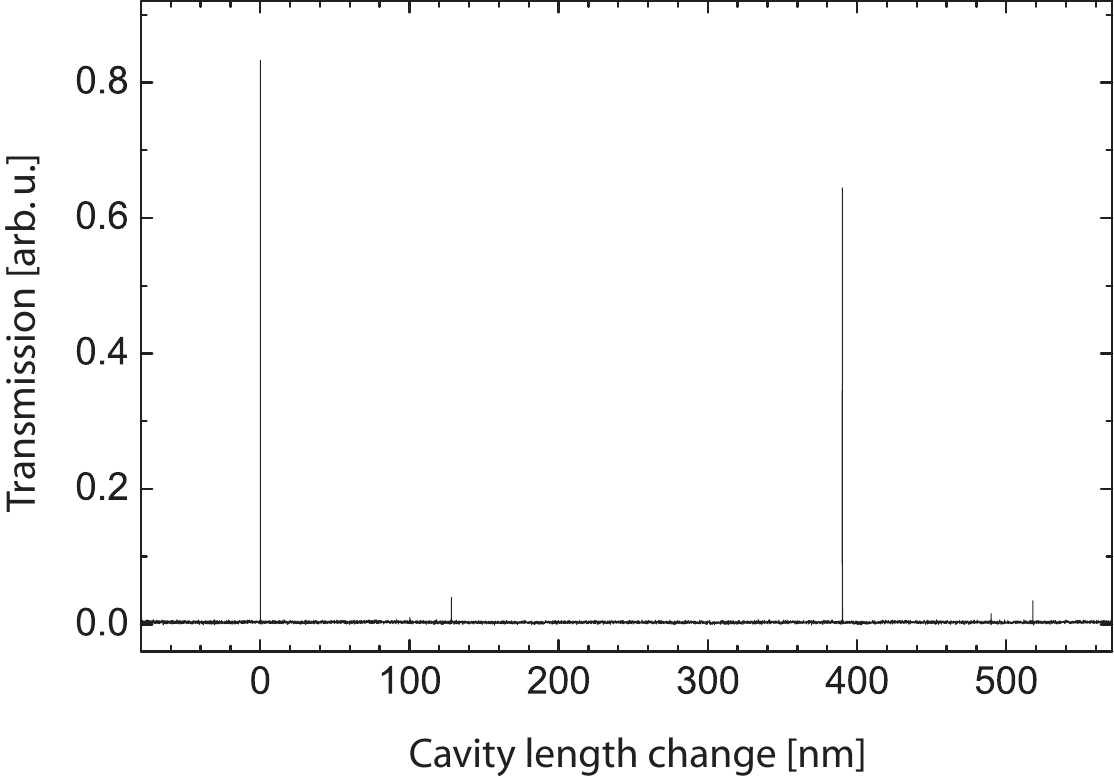}
  \includegraphics[width=0.49\textwidth]{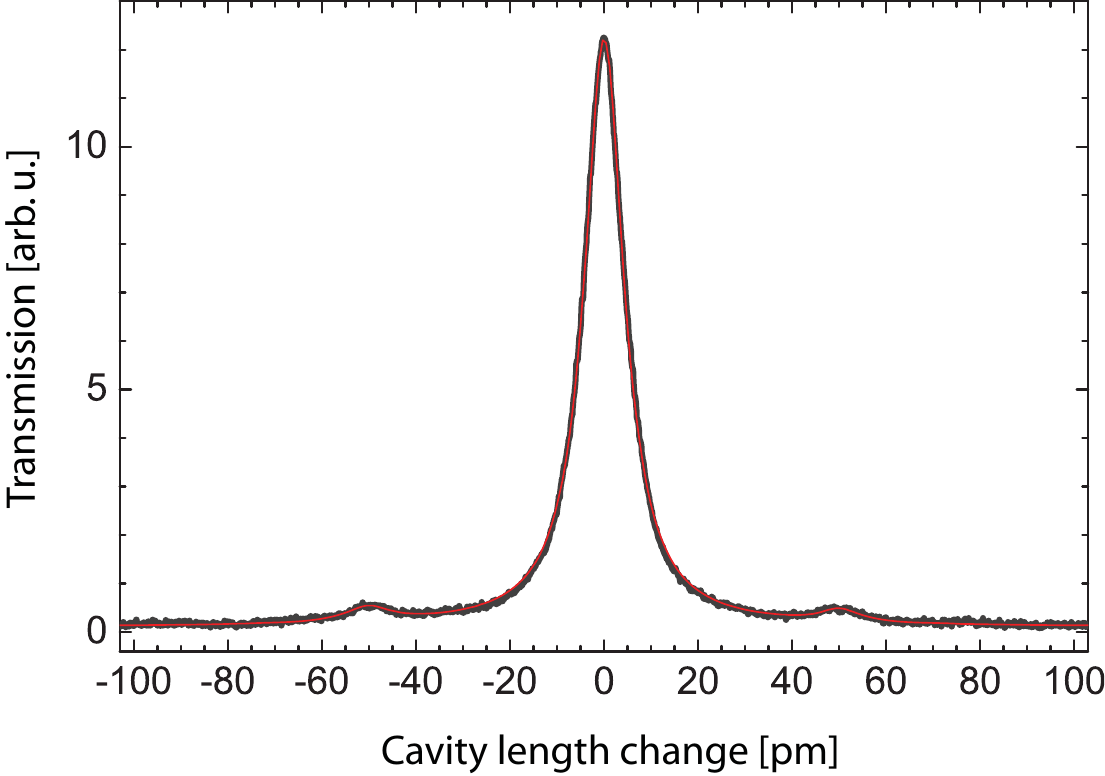}
  \caption{\textit{Left:} Transmission of the FFP1 cavity at
    780\,nm. Cavity length is scanned with a piezo actuator, which has
    been calibrated to obtain an accurate length scale. With optimized
    alignment, the TEM$_{00}$ mode is more than 20 times stronger than
    all higher order modes. \textit{Right:} Linewidth of the same
    cavity, measured in transmission. The laser is modulated with
    sidebands at $\pm 500\,$MHz for frequency calibration. The
    measurement yields a FWHM linewidth of
    100.4\,MHz. \label{fig:cavScan}}
\end{figure}

As a first method to characterize different FFPCs over the
whole usable cavity length, we measure transmission spectra at fixed
laser frequency, using a piezo actuator to scan $L$
(fig.~\ref{fig:cavScan}). This can be done with both the test mount
and the miniature mount. To obtain a calibration of the length scale,
we have used various methods to simultaneously couple several laser
frequencies with a well-known difference. To obtain a difference in
the GHz range, we use two lasers, typically $\lambda_1=780.241\,$nm,
locked on the $^{87}$Rb D2 line, $\lambda_2=780.320\,$nm measured with
a 6 digit Burleigh wavemeter, $c/\lambda_1-c/\lambda_2\approx
38.9\,$GHz). With this wavelength calibration, the FSR and the
distance $L=c/2 \FSR$ can be determined with an uncertainty below
$500\,$nm.  To measure the cavity linewidth, a smaller frequency
difference is useful, and can be obtained by modulating a single laser
at a known frequency in the RF range (fig.~\ref{fig:cavScan},
left). $\Fi$ is obtained as the ratio of the FSR and linewidth results.

\begin{figure}[tb]
  \centering
  \includegraphics[width=0.4\textwidth]{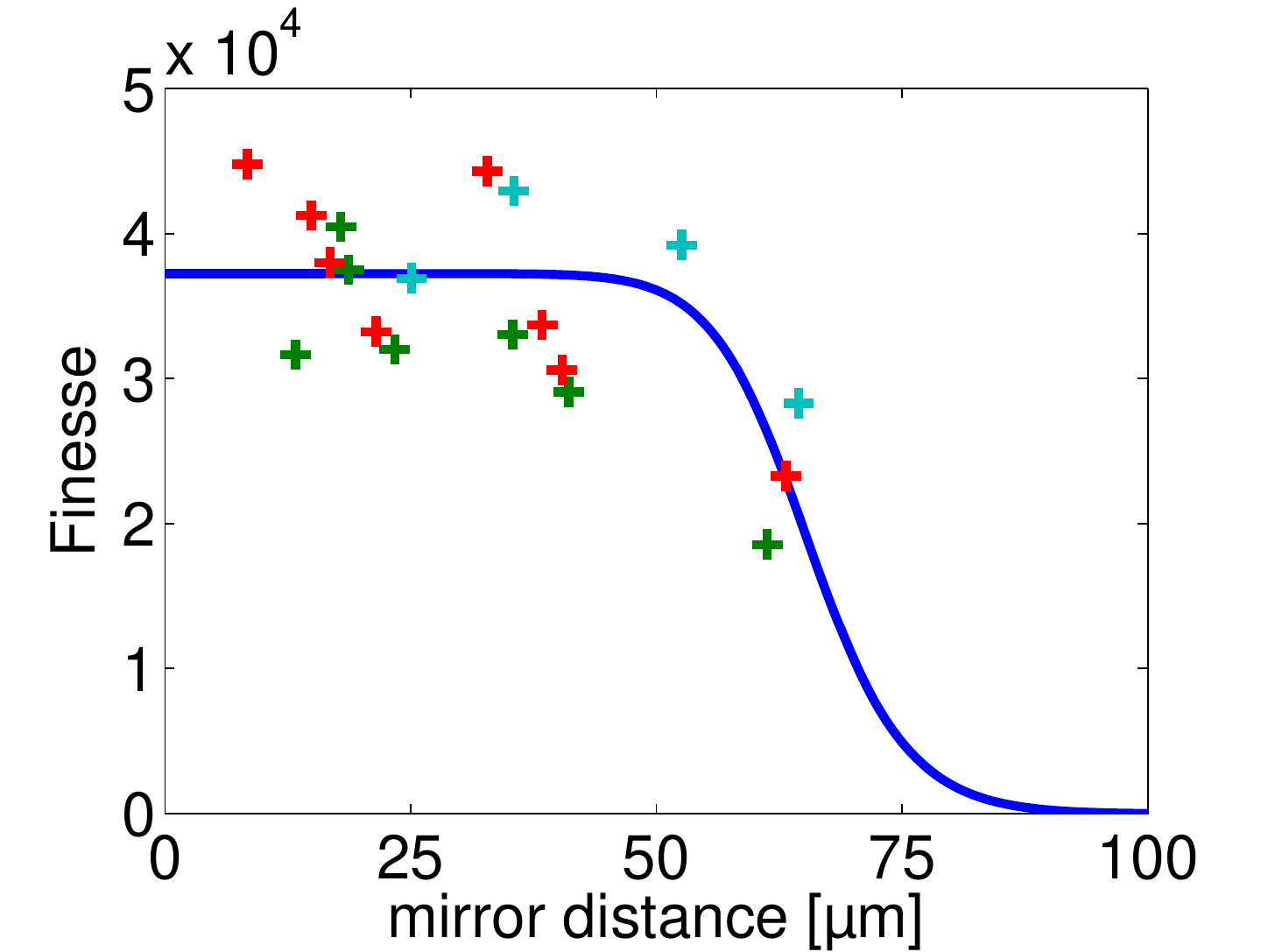}
  \caption{Finesse
    measurement of three different cavities. For small
    distance,  $\Fi=37000\pm5000$.  The solid curve is the calculated finesse
    including clipping loss for $R_1=350\,\mu$m, $R_2=100\,\mu$m and
    $D_2=23\,\mu$m. \label{fig:FvsL}}
\end{figure}

For a set of three different fiber pairs, we have performed finesse
measurements over the whole length range of the cavities
(figure~\ref{fig:FvsL}). All cavities had SM fibers on the input side
and MM fibers at the output, and ROC was large on the
SM side and small on the MM side: $R_1$ between 300 and $400\,\mu$m,
$R_2$ between 60 and $160\,\mu$m, $D_1$ between 20 and $40\,\mu$m, and
$D_2$ between 30 and $40\,\mu$m. For small mirror distance, we obtain
a Finesse of $\Fi=FSR/\delta\nu=37000\pm5000$ . The large
scatter comes from variation in the alignment quality and from the
limited mechanical stability of the test mount. The finesse drops
significantly for $L>50\,\mu$m and the resonances vanish in the noise
level above $L\sim70\,\mu$m. This behaviour is well reproduced by
including clipping loss (eq.~\ref{eq:cliploss})
into the calculated finesse: The solid curve in fig.~\ref{fig:FvsL} is
calculated for $R_1=350\,\mu$m, $R_2=100\,\mu$m and
$D_2=23\,\mu$m. Note that the finesse drops sharply over a small
distance range. The ``cutoff'' length at which this occurs depends
strongly on the effective mirror diameters. For the cavities in
fig.~\ref{fig:FvsL}, where the waist is located close to mirror 1, it
is dominated by $D_2$. It will be interesting to repeat such
$\Fi(L)$ measurements for mirrors with different diameters, as this will
lead to more accurate predictions of the $D$ values required to reach
a desired cavity length.

For cavities FFP1 and FFP2, mounted in the miniature mount, we used an
improved version of the above protocol. Both lasers were locked to Rb
resonances, with a frequency difference of 212\,MHz. In this setup,
the length could be determined to better than $\lambda/2$ so that the
resonance order is exactly known: $d_1=99\lambda/2=38.62\,\mu$m,
$d_2=69\lambda/2=26.9\,\mu$m. With a frequency calibration via
amplitude modulation of the laser, we obtain a value for the resonance
linewidth of $\delta\nu_1=100.4$MHz and $\delta\nu_2=156$MHz
(fig.~\ref{fig:cavScan}). Combining these measurements yields
$\Fi_1=38600$, $\Fi_2=36000$.

For the FFP1 cavity, we have also measured the dependence on the
polarization of the incoming beam. We observe a splitting of
$200\,$MHz between two orthogonal, linear input polarizations. Similar
birefringence has been observed in macroscopic high-finesse
cavities. In our case, it may be related to the ellipticity observed
on some of the mirrors.

\subsection{Total transmission and mode matching}
\label{sec:couplingExp}

We have measured transmitted and reflected powers for several cavities
in the test and miniature mounts.
Our main goal was to obtain information about the coupling efficiency
$\mmeff$ between a SM fiber and the cavity. We have investigated
this question with cavities having a SM fiber at the input and a MM
fiber on the output side. The on-resonance total transmission (from
the free-space beam before the first fiber to the beam leaving the
second fiber) is
\begin{equation}
  \T_{\textrm{tot}}=\mmeff_2\left(\frac{\T}{\T+\L}\right)^2\mmeff_1\fceff \,.
\end{equation}
The analysis is more complicated than for macroscopic FPCs because
neither the coupling efficiency from the free-space beam into the
incoupling fiber, $\fceff$, nor the mode-matching efficiencies between
the fibers and the resonator mode, $\mmeff_{1,2}$, can be determined
directly in a non-destructive way. (It would be possible to measure
$\fceff$ by breaking the fiber.) The measured intensity transmission
from before the input of the SM input fiber to after the output of the
MM output fiber is 0.094 on resonance for FFP1. (Similar results were
obtained for other cavities.) From the mirror transmission measured on the macroscopic substrates,
$\T=31\,$ppm, and $\T+\L=\pi/\Fi=85\,$ppm, the on-resonance transmission of
the perfectly mode-matched cavity would be $\left
  (\T/\T+\L\right)^2=0.13$. (Note again that this value is limited by
the sub-optimum coatings.) We infer that $\mmeff_2 \mmeff_1
\fceff=0.094/0.13=0.72$. A lower bound for $\mmeff_{1}$ can be
obtained by attributing all this loss to the input mode-matching, so
$\mmeff_{1}\ge =0.72$. In reality, $\fceff$ is responsible for a
significant part of the losses. We have estimated $\fceff$ by
replacing the cavity input fiber with an ``open'' fiber (no mirror on
the output side) of the same type. In this case, the fiber coupling
efficiency is easily measured, and never exceeded
$\fceff_{1,\max}=0.85$ in our setup. A more realistic value is therefore
$\mmeff_{1}\gtrsim 0.72/0.85 =0.85$. The theoretical maximum value for perfect alignment (eq.~\ref{eq:mmFull}) is $\mmeff_{1}=0.979$.

Some additional information can be obtained from a reflection
measurement. Let us call $P_i$ the power incident on the fiber input
port and $P_r$ the power that emanates from the same port in the
reverse direction (which we can measure using a beam splitter). Off
resonance or with the second fiber removed, we have 
\begin{equation}
P_r = \reff R \fceff P_i\approx \reff\fceff P_i\,.
\end{equation}
Here we assume that all light from the input beam which is not coupled
into the fiber is lost and contributes neglegibly to the reflected
beam. The parameter $\reff<1$ takes into account that some of the
light reflected back from the mirror is not guided by the fiber
because the mirror is not planar and may be misaligned\footnote{In
  the experiment, this light can be observed as a diffuse glowing of
  the mirrored fiber end.}. (Note that in general $\reff\neq\mmeff_1$
  because the reflected mode is not identical to the cavity mode.)  A
measurement of $P_r$ yields the product $\reff\fceff$. For FFP1, we
obtained $\reff\fceff=0.68$. Measuring $\fceff$ with an open fiber
gave $\fceff\approx 0.85$ in this case, so we obtain
$\reff\approx 0.8$. For various test cavities, we obtained
$\reff=0.7\ldots0.85$. 

On resonance, a symmetric cavity excited by a perfectly mode-matched
power $P_{i,0}$ reflects
\begin{equation}
P_{r,0} = P_{i,0} \, 
R \frac{\L^2}{(\T+\L)^2}\approx P_{i,0}\frac{\L^2}{(\T+\L)^2} \,.
\label{eq:mmref}
\end{equation}
Again, we cannot measure $P_{i,0}$ and $P_{r,0}$ directly. $P_r$ has
two contributions \cite{Hood01} on resonance: the cavity, which is now
excited by the mode-matched power $P_{i,0}=\mmeff\fceff P_i$, reflects
according to eq.~\ref{eq:mmref}. The mode-matched part of this
reflection, $\mmeff P_{r,0}$, is guided back by the fiber. Second, the
non-modematched component of the input light, $(1-\mmeff)\fceff P_i$
is reflected on the mirror and a fraction $\reff'$ of it guided
back. Unfortunately, $\reff'\neq\reff$ because the two constants
relate to different incident modes. (We expect $\reff'<\reff$.) In
total, we have

\begin{equation} 
P_r = P_i\left((1-\mmeff)\reff'\fceff  + \mmeff^2\fceff
          \frac{\L^2}{(\T+\L)^2} \right)
\label{eq:PrRes}
\end{equation}
Because of the additional unknown constant $\reff'$, this formula is
difficult to exploit.

\subsection{Mode geometry and cavity QED parameters}

Strong coupling between an emitter and a resonant cavity gives rise to
a split resonance, displaced by $\pm g$ with respect to the empty
cavity resonance, where $g$ is the coherent coupling rate (``vacuum
Rabi splitting''). Measuring the resonance frequencies of the coupled
system can thus be used to determine $g$.  The coupling rate $g_0$ in
sec.~\ref{sec:CQEDparams} is the maximum coupling calculated for a
point-like particle localized in an antinode of the cavity field.  It
is straightforward to extend to the position-dependent coupling
$g(\mathbf{r})$. Furthermore, for $N$ independent atoms with a
density distribution $\rho(\mathbf{r}))$, one finds:
\begin{equation}
  g_N=\sqrt{N}\overline{g}_1\,,\:\textrm{with}\:\:
  \overline{g}_1^{\,2}=\int\frac{\rho(\mathbf{r})}{N} 
  |g(\mathbf{r})|^2 d\mathbf{r} \,.
\end{equation}
In \cite{Colombe07}, we have measured $g_N$ for Bose-Einstein
condensates with variable $N$. The atoms were strongly confined and well
localized in the region of maximum coupling of cavity FFP1 described
above. $N$ was determined independently by absorption imaging. An
estimated uncertainty of about a factor 2 in the $N$ measurement
dominates the precision to which we can determine $g_0$ from this
measurement. $\rho(\mathbf{r})$ depends on $N$ and could in principle
be taken into account precisely. However, within the uncertainty
imposed by the $N$ measurement, the distribution can be considered
point-like for $N\lesssim 1000$. $g_0$ can be estimated simply as the
prefactor of a $\sqrt{N}$ function fitted to the data for small
$N$. This estimate gives $g_0\approx 200\,$MHz, in agreement with the
calculated value.

\section{Optomechanical bistability}
\label{sec:bistability}
\begin{figure}[tb]
  \centering
  \includegraphics[width=0.49\textwidth]{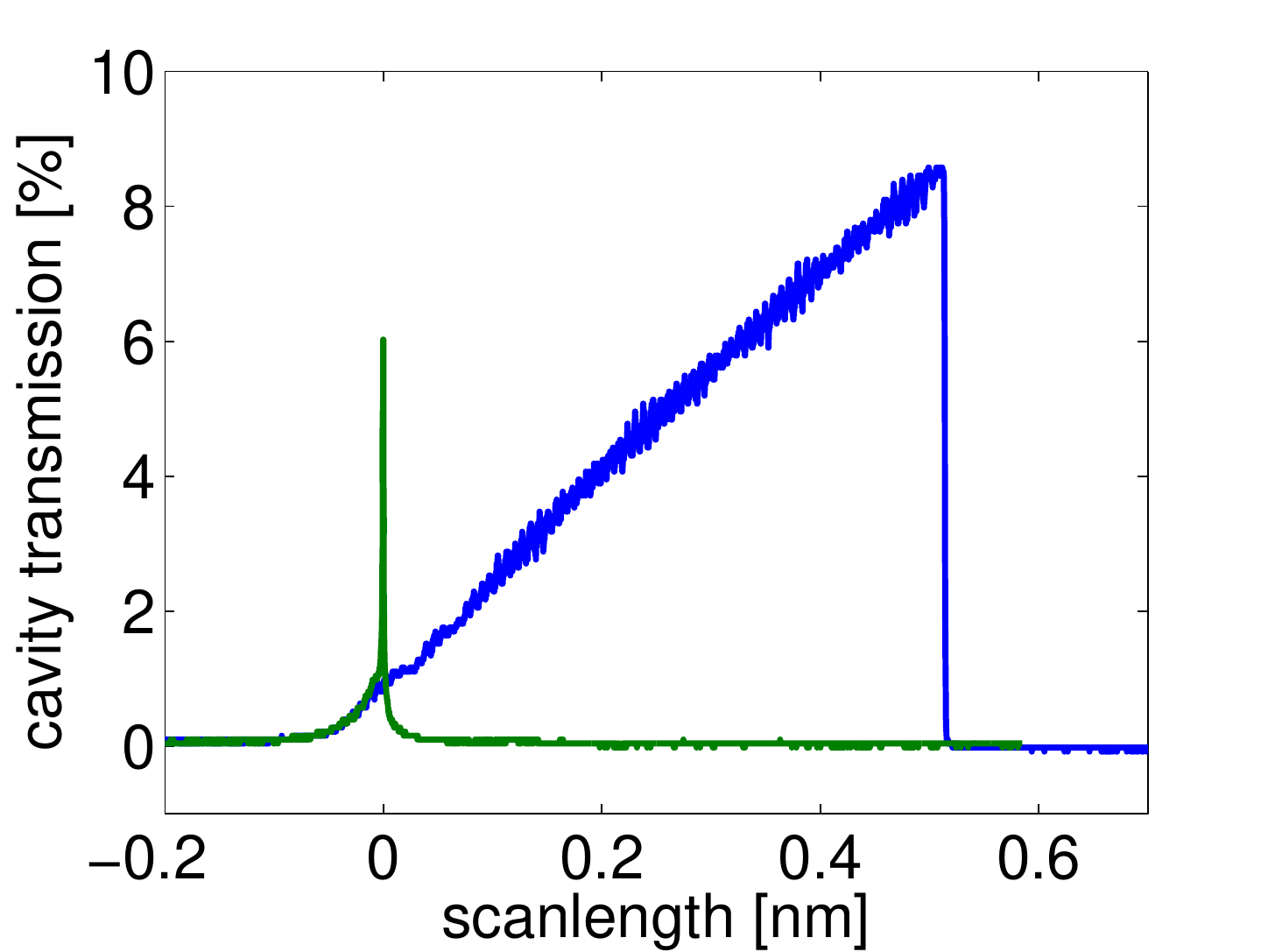}
  \includegraphics[width=0.49\textwidth]{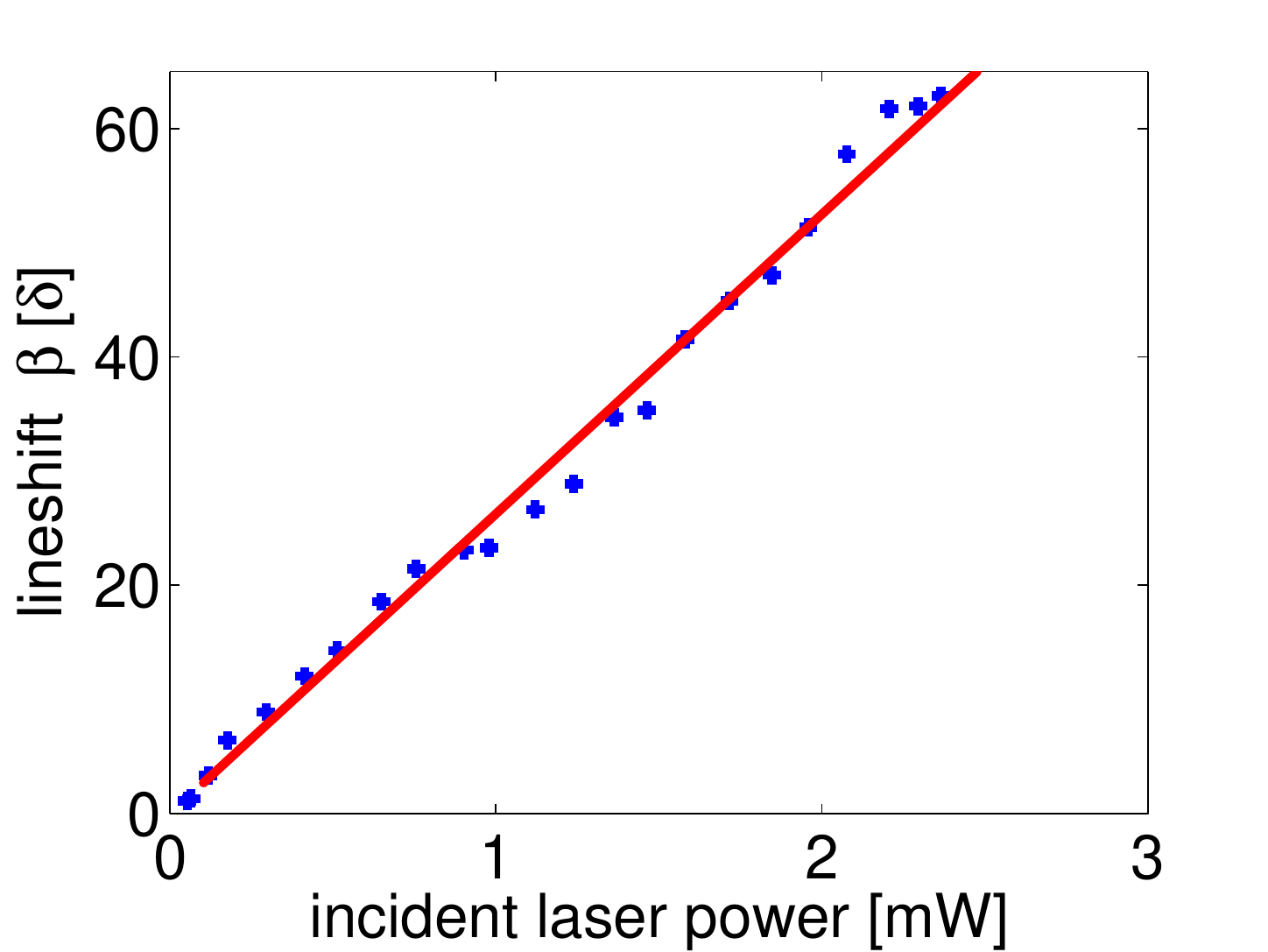}
  \caption{Optical bistability by absorption in the mirrors: (a)
    Absorption induced thermal expansion leads to two different cavity
    line shapes when increasing / decreasing the cavity
    length. Measurement taken with a scan speed
    $\nu_{\textrm{scan}}=2\times10^6~$1/s and $P_{i,0}=2.3~$mW. (b) The
    lineshift $\beta$ as a function of incident power for
    $\nu_{\textrm{scan}}=2\times10^5~$1/s. A linear fit to the data
    yields $\beta/P_{i,0}=26.3~$1/mW.}
  \label{fig:opticalbistability}
\end{figure}

When scanning the cavity length using an optical power above a certain
threshold, we find strong optical bistability induced by absorption in
the mirror: the small fraction of the intracavity power that is
absorbed in the mirror coating causes local heating and thermal
expansion of the mirror and the fiber. This changes the effective
cavity length and therefore the cavity field, leading to
bistability. We can observe strongly broadened resonance lines when
the cavity is scanned towards increasing length, showing an almost
linear increase of transmitted power followed by a rapid drop. This is
explained by the expansion of the substrate compensating the length
change of the scanning piezo. In the other direction (towards shorter
cavity length), we see a narrowed line, corresponding to the cavity
being pushed across the resonance by the expansion.
Figure~\ref{fig:opticalbistability}(a) shows the characteristic
lineshape for both scan directions.

From the width of the broadened resonances, the absorption in the
mirror coatings can be inferred. We measure the total width, as
defined in \cite{An97}, and express it as a multiple $\beta$ of the
unperturbed (low-power) linewidth $\kappa$ (i.e., the total width is
$\beta\kappa$).  In the adiabatic limit, where the temperature
distribution in the coating and fiber reaches steady state, $\beta$ is
determined by \cite{An97}

\begin{equation}\label{eq:lineshift}
  \beta=\frac{8c_0C_{\mathrm{ex}}P_{i,0}\T\A\,\Fi^3}{\pi^3\lambda k}
\end{equation}

with $C_{\mathrm{ex}}$ the effective thermal expansion coefficient of
the coating and the fiber, $c_0 \approx 3.2$ a geometrical factor,
$P_{i,0}$ the incident, modematched laser power in the fiber, and
$k=1.4\,\textrm{W}\textrm{m}^{-1}\textrm{K}^{-1}$ the thermal
conductivity of the fiber.  Adiabaticity is met for slow scan velocity
$\nu_{\textrm{scan}}<\tau_r^{-1}$, where
$\nu_{\textrm{scan}}=\dot{\omega}_\textrm{cav}/\kappa$ is the scan
velocity in units of the cavity linewidth and $\tau_r=c_0sw_0^2/k$ the
thermal diffusion time with the specific heat per unit volume $s =
1.5\times10^6\,\textrm{J}\textrm{m}^{-3}\textrm{K}^{-1}$ for SiO$_2$.
For our parameters, $\tau_r = 25~\mu$s and steady state conditions are
expected for $\nu_{\textrm{scan}} < 4\times10^4\,\textrm{s}^{-1}$. Due
to the limited mechanical stability of the test mount, we had to use a
slightly faster scan speed
$\nu_{\textrm{scan}}=2\times10^5\,\textrm{s}^{-1}$, so that our measured
$\beta$ values are a lower bound to the steady-state value.
Fig.~\ref{fig:opticalbistability}(b) shows the measured $\beta$ as a
function of the incident power. We obtain $\beta/P_i =
26.3\,$mW$^{-1}$.  To deduce the mirror absorption from this
measurement, the contribution of the coating to the effective thermal
expansion coefficient has to be known. From a finite-element
simulation, we find that the transient temperature profile extends
$\sim200\,\mu$m into the fiber, and that the mirror coating
contributes $\sim10\%$ to the expansion. Hence $C_\mathrm{ex}$ is
dominated by the value for SiO$_2$
$C_\mathrm{ex,SiO_2}=0.55\times10^{-6}\,$K$^{-1}$ with a $\sim5\%$
contribution of the Ta$_2$O$_5$ layers with
$C_\mathrm{ex,Ta_2O_5}=3.6\times10^{-6}\,$K$^{-1}$.  With the measured
values for the finesse and the mirror transmission, equation
(\ref{eq:lineshift}) yields an absorption loss $\A=35\pm 10\,$ppm, in
good agreement with the value obtained in
Sec.~\ref{sec:coatingProps}. The error for this value stems from the
uncertainty of the steady-state value of $\beta$ and of the
Ta$_2$O$_5$ contribution to the thermal expansion. A more accurate
determination could be obtained by measuring the temporal response of
the thermal expansion \cite{An97}.  To avoid the bistability effect,
intracavity power should be limited to
$P_{i,0}\mathcal{T}/(\mathcal{T}+\mathcal{L})^2<3\,\mathrm{W}
\times\mathrm{ppm}/\mathcal{A}$.

\section{Conclusion}

As the results show, FFPCs with laser-machined mirrors combine several
desirable properties in a single device. The most important of these
are a small mode waist, high finesse, efficient and robust fiber
coupling, and open access to the cavity mode. While each of these
features individually can also be realized with other cavity types,
their combination in a single device is unique to our knowledge, and
explains the remarkable interest they are meeting since the
publication of \cite{Colombe07}. We expect them to be useful in a
correspondingly wide range of applications. The devices we have
realized so far do not exploit yet the full potential of the
laser-machined fiber surface: they are limited by the scatter and
absorption losses of sub-optimum mirror coatings. An obvious next step
is therefore to have the same fibers coated with ``supermirror''
\cite{Hood01} coatings, which should enable simultaneous significant
increase of finesse and overall transmission. Applications that we are
currently investigating in collaboration with specialists from various
domains include strong-coupling cavity QED with trapped ions, 
cavity optomechanics, and coupling to solid-state emitters such as
quantum dots and color centers in diamond.

\section*{Acknowledgements}
We thank Richard Warburton for fruitful discussions on fiber mirror
fabrication, Jean Hare and Fedja Orucevic for kindly giving us access
to their CO$_2$ laser setup at LKB, CeNS (Munich) for access to their
cleanroom, INSP (Paris), ESPCI (Paris) and Didier Chatenay's group
at the ENS Physics Department for access to their AFM and
interferometric microscopes, and Stephan G\"unster of LZH and his team
for the mirror coating.

We gratefully acknowledge financial support for this work from a EURYI
award and the SCALA Integrated Project of the EU.

\section*{Appendix: Mismatch between fiber and cavity modes}

In the following we calculate the coupling efficiency between the mode
of a SM fiber and the cavity mode. Both modes are assumed to be
gaussian. We take into account the lensing effect due to the concave
fiber mirror, and we include the mismatch of wavefront
curvatures. What is still neglected in this calculation is the finite
coating thickness and any misalignment (centering error) between the
mirror and the fiber core.

For concreteness, let us consider light in a SM fiber incident from
the left onto the first mirror of an FFPC. The fiber mode 
is characterized by its mode field radius $w_f$. The radius of
curvature of the mirror is $R$. We also know the mode radius
$w_m$ of the cavity mode on this mirror.

The power transmission coefficient between two gaussian modes is \cite{Joyce84}
\begin{equation}
\epsilon=
  \frac{4}{\left(\frac{w_0'}{w_0}+\frac{w_0}{w_0'}\right)^2 +
           \frac{s^2}{z_R z_R'}}
\label{eq:mmFullLit}
\end{equation}
where $s=z_0-z_0'$ is the distance between the mode waist positions.
In our case, the first mode in this formula is the one leaving the
fiber through the concave mirror, which has undergone diffraction as
the mirror acts like a plano-concave lens. As the mirror is very thin,
the mode after passing through the mirror still has the radius $w_f$,
but has wavefront curvature
\begin{equation}
  R_1=\frac{R}{n_f-1}\,,
\end{equation}
where $n_f$ is the index of the fiber (we can neglect the very small
difference between the core and cladding index).

Knowing its radius and curvature at the mirror position, the first
mode is fully determined. The second mode in eq.~(\ref{eq:mmFullLit}) is
the cavity mode, which is fully determined by $R$ and
$w_m$. Deriving $w_0$, $w_0'$, $z_R$ and $z_R'$ and inserting into
eq.~(\ref{eq:mmFullLit}) gives the final result of eq.~\ref{eq:mmFull}.

Neglecting the lensing effect of the mirror and the mismatch of
wavefront curvature amounts to assuming
\begin{equation}
\left(\frac{\pi n_f w_f w_m}{\lambda R}\right)^2
\ll
\left(\frac{w_f}{w_m}+\frac{w_m}{w_f}\right)^2\,,
\end{equation}
which is a good approximation in many cases, as shown in
fig.~\ref{fig:fceff}.

\section*{References}
\bibliographystyle{njpph2}

\begin{thebibliography}{10}

\bibitem{Haroche06}
Haroche S and Raimond J~M {\it Exploring the Quantum\/} (Oxford: Oxford
  University Press, 2006)

\bibitem{Mabuchi02}
Mabuchi H and Doherty A~C 2002 {\it Science\/} {\bf 298} 1372

\bibitem{Zoller05}
Zoller P~e~a 2005 {\it Eur. Phys. J. D\/} {\bf 36} 203

\bibitem{Kimble08}
Kimble H~J 2008 {\it Nature\/} {\bf 453} 1023

\bibitem{Banaszek09}
Banaszek K, Demkowicz-Dobrza{\'n}ski R and Walmsley I~A 2009 {\it Nature
  Photonics\/} {\bf 3} 673

\bibitem{Leroux10}
Leroux I~D, Schleier-Smith M~H and Vuleti{\'c} V 2010 {\it Phys.~Rev.~Lett.\/}
  {\bf 104} 073602

\bibitem{Law97}
Law C~K and Kimble H~J 1997 {\it J. Mod. Opt.\/} {\bf 44} 2067

\bibitem{Vahala03}
Vahala K~J 2003 {\it Nature\/} {\bf 424} 839

\bibitem{Colombe07}
Colombe Y, Steinmetz T, Dubois G, Linke F, Hunger D and Reichel J 2007 {\it
  Nature\/} {\bf 450} 272

\bibitem{Luo09}
Luo L, Hayes D, Manning T, Matsukevich D, Maunz P, Olmschenk S, Sterk J and
  Monroe C 2009 {\it Fortschr.~Phys.\/} {\bf 57} 1133

\bibitem{Jelezko06}
Jelezko F and Wrachtrup J 2006 {\it Phys. Stat. Sol. A\/} {\bf 203} 3207

\bibitem{Deveaud06}
Deveaud B, ed. {\it The Physics of Semiconductor Microcavities\/} (Weinheim:
  Wiley-VCH, 2006)

\bibitem{Shields07}
Shields A~J 2007 {\it Nature Photonics\/} {\bf 1} 215

\bibitem{Jayich08}
Jayich A~M, Sankey J~C, Zwickl B~M, Yang C, Thompson J~D, Girvin S~M, Clerk
  A~A, Marquardt F and Harris J~G~E 2008 {\it New Journal of Physics\/} {\bf
  10} 095008

\bibitem{Favero09}
Favero I, Stapfner S, Hunger D, Paulitschke P, Reichel J, Lorenz H, Weig E~M
  and Karrai K 2009 {\it Optics Express\/} {\bf 17} 12813

\bibitem{Lounis05}
Lounis B and Orrit M 2005 {\it Rep.~Prog.~Phys.\/} {\bf 68} 1129

\bibitem{Poldy08}
Poldy R, Buchler B~C and Close J~D 2008 {\it Phys.~Rev.~A\/} {\bf 78} 013640

\bibitem{Salik02}
D. S~M, Nicolas C, Carr{\'e} A and Carracci S~J in {\it Proc. 28th European
  Conf. on Optical Communication, Copenhagen, Denmark\/} (2002)  Paper 10.4.6

\bibitem{Steinmetz06}
Steinmetz T, Colombe Y, Hunger D, H{\"a}nsch T~W, Balocchi A, Warburton R~J and
  Reichel J 2006 {\it Appl.~Phys.~Lett.\/} {\bf 89} 111110

\bibitem{Trupke05}
Trupke M, Hinds E~A, Eriksson S, Curtis E~A, Moktadir Z, Kukharenka E and Kraft
  M 2005 {\it Appl.~Phys.~Lett.\/} {\bf 87} 211106

\bibitem{Muller09}
Muller A, Flagg E~B, Metcalfe M, Lawall J and Solomon G~S 2009 {\it
  Appl.~Phys.~Lett.\/} {\bf 95} 173101

\bibitem{Hunger08b}
Hunger D, Deutsch C, Warburton R and Reichel J 2010   to be published

\bibitem{Vernooy98}
Vernooy D~W, Furusawa A, Georgiades N~P, Ilchenko V~S and Kimble H~J 1998 {\it
  Phys.~Rev.~A\/} {\bf 57} R2293

\bibitem{Armani03}
Armani D~K, Kippenberg T~J, Spillane S~M and Vahala K~J 2003 {\it Nature\/}
  {\bf 421} 925

\bibitem{Bennett92}
Bennett J~M 1992 {\it Meas. Sci. Technol.\/} {\bf 3} 1119

\bibitem{Hood01}
Hood C~J, Kimble H~J and Ye J 2001 {\it Phys.~Rev.~A\/} {\bf 64} 033804

\bibitem{Siegman86}
Siegman A~E {\it Lasers\/} (Mill Valley: University Science Books, 1986)

\bibitem{Reitzenstein07}
Reitzenstein S, Hofmann C, Gorbunov A, Strau\ss\ M, Kwon S~H, Schneider C,
  L\"{o}ffler A, H\"{o}fling S, Kamp M and Forchel A 2007 {\it Appl. Phys.
  Lett.\/} {\bf 90} 251109

\bibitem{vanEnk01}
van Enk S~J and Kimble H~J 2001 {\it Phys.~Rev.~A\/} {\bf 63} 023809

\bibitem{Thompson08}
Thompson J~D, Zwickl B~M, Jayich A~M, Marquardt F, Girvin S~M and Harris J~G~E
  2008 {\it Nature\/} {\bf 452} 72

\bibitem{An95}
An K, Yang C, Dasari R and Feld M 1995 {\it Opt.~Lett.\/} {\bf 20} 1068

\bibitem{An97}
An K, Sones C, Fang-Yen R, Dasari R and Feld M 1997 {\it Opt.~Lett.\/} {\bf 22}
  1433

\bibitem{Joyce84}
Joyce W~B and DeLoach B 1984 {\it Appl. Opt.\/} {\bf 23} 4187

\end{thebibliography}

\end{document}